\documentclass[preprint,showpacs]{revtex4-1}
\usepackage{amsmath}
\usepackage{graphicx}
\usepackage{latexsym}
\newcommand{\ba}{\begin{eqnarray}}
\newcommand{\ea}{\end{eqnarray}}

\newcommand{\es}{&=&}

\newcommand{\nn}{\nonumber \\}
\usepackage{hyperref}

\newcommand{\bs}{\boldsymbol} 
\newcommand{\T}{\mathcal{T}}
\usepackage{color}
\begin{document}
\title{\bf\large{Wigner Distributions For Gluons}}

\author{\bf Jai More$^{a}$, Asmita Mukherjee$^{a}$ and Sreeraj Nair$^{b}$}
 
\affiliation{$^{a}$  Department of Physics,
 Indian Institute of Technology Bombay,\\ Powai, Mumbai 400076,
 India.\\  
 $^{b}$ Department of Physics, Indian Institute of Science Education and Research Bhopal\\
 Bhopal Bypass Road, Bhauri, Bhopal 462066 India}

\date{\today}
\begin{abstract}
We investigate the gluon Wigner distributions for unpolarized, longitudinally polarized and transversely polarized target state. Instead of a nucleon,  we take the target state to be a quark dressed with a gluon at one loop and investigate the gluon Wigner distributions at leading twist.  Better numerical convergence is obtained compared to an earlier study, that removes the regulator dependence of the results.  We present a first calculation of the Wigner distribution for the transversely polarized target and linearly polarized gluon. We study the spin densities in momentum and impact parameter space. We also investigate the quark and gluon helicity and orbital angular momentum distributions  at small-$x$.

\end{abstract}
\maketitle


\section{Introduction}
Nucleon spin puzzle has been a provocating issue since a couple of decades and the goal is to study, 
explore and understand the three dimensional 
picture of hadrons in terms of quarks and gluons. The generalized parton correlation functions (GPCFs)\cite{Meissner09} provide a framework to study the composite structure of hadrons. GPCFs are fully unintegrated off diagonal quark-quark correlator which are functions of 4-momentum of quark and 4-momentum  $\Delta$, which is transferred by the hadrons
and contains a wealth of information of the nucleon. The GPCFs are also related to  
quantum mechanical analogue of classical phase space distribution called as
Wigner distributions \cite{Wigner32, Ji03, Belitsky04}. On integrating the GPCFs over the light front energy $(k^-)$ one obtains generalized transverse- momentum-dependent distributions (GTMDs). GTMDs  and the Wigner distributions are Fourier transforms of each other \cite{Meissner09, Burkardt16}. The matrix element of Wigner distributions are six dimensional and are functions of three position and three momentum variables \cite{Ji03, Belitsky04}. It is convenient to use a  five-dimensional Wigner distribution as a function of two transverse positions and three momentum coordinates in light-front formalism \cite{Lorce11}.
These are often considered as mother distributions of the transverse-momentum dependent distribution (TMDs)\cite{ Mulders01, Ji05, Goeke05} and 
generalized parton distributions (GPDs) \cite{Ji97,Goeke01, Diehl02, Diehl03, Ji04, Belitsky05, Boffi07}.   
GPDs are obtained by integrating Wigner distribution over transverse momentum ${\bs k _\perp}$ while TMDs are obtained by integrating the Wigner distributions over the transverse position ${\bs {\bs b _\perp}}$.

Wigner distributions  are not always positive definite.  So far the Wigner distributions of quarks and gluon have only been studied through model calculations, for example in constituent quark model and chiral quark soliton model \cite{Lorce11}, spectator model \cite{Liu15} and holographic model \cite{Chakrabarti16, Chakrabarti17}. Wigner distributions of nucleons in a nucleus were studied in \cite{Yuri16}.  Very recently the possibility to access the quark GTMDs in an exclusive double Drell-Yan process has been investigated in \cite{Bhattacharya17}. Most of the models have no gluonic degrees of freedom, and so  gluon GTMDs and gluon Wigner distributions cannot be investigated in such models. 
Although interest in quark GTMDs are rather recent, gluon GTMDs and unintegrated gluon correlators have been introduced and discussed in the context of small $x$ physics for diffractive vector meson production \cite{Martin00} and for Higgs production at Tevatron and LHC \cite{Khoze00} quite some time back. The details of operator definitions, as well as parametrizations of gluon GTMDs, are given in \cite{Lorce13}. There is also a growing interest to understand how to probe the gluon GTMDs experimentally that are related to the gluon Wigner distributions by a Fourier transform.  In \cite{Hatta16} the authors proposed to study small $x$ gluon GTMDs in diffractive dijet production, and also in \cite{HattaXiao16} for correlated hard diffractive dijet production in deep inelastic scattering.   Exclusive dijets in deep inelastic scattering within the formalism of color glass condensate was first studied in \cite{Tolga16}. Elliptic gluon GTMDs were studied in \cite{Zhou16}. In \cite{Hatta17} the authors proposed to probe the gluon   GTMDs and Wigner distributions in DVCS at small $x$. In \cite{Hagiwara16} gluon Wigner and Husimi distributions were investigated. Husimi distributions contain a Gaussian factor in the integrand that makes them positive definite, unlike Wigner distributions that are not positive definite.  The disadvantage is that upon integration, the Husimi distributions do not reduce to known TMDs \cite{Hagiwara17} suggested to probe the gluon Wigner distributions in ultra-peripheral $pA$ collision.  

Compared to the quark GTMDs,  gluon GTMDs   and Wigner distributions have more complex process dependence due to the presence of two gauge links in the gluon correlator required by color gauge invariance \cite{Bomhof06,Bomhof08,Buffing13},  the simplest combinations are $++$  
(Weizsacker-Williams or WW-type) and $+-$ (dipole type). Recent interest in quark and gluon Wigner distributions and GTMDs is related to the fact that these can provide important information about the  so far unknown quark and gluon orbital angular momentum (OAM) in the nucleon \cite{Hagler04, Lorce12, Hatta12,Burkardt13}.  It was shown in \cite{Hatta16} that both the WW type and dipole type gluon GTMDs give the same gluon OAM distribution.  In models of the nucleon without gluonic  degrees of freedom, one cannot study the gluon Wigner distributions/GTMDs. The possibility to probe the gluon Wigner distribution and gluon orbital angular momentum at the future electron-ion collider was studied in \cite{Ji17}.  In \cite{ Mukherjee14, Mukherjee15} we investigated the quark and gluon Wigner distributions and OAM for a simple composite spin-${1\over 2}$ target state, namely for a  quark dressed with a gluon at one loop in QCD. This may be considered as a field theory inspired perturbative model having a gluonic degree of freedom.  The quark has non-zero mass and both quark and gluon have non-zero transverse momenta.  Wigner distributions were expressed in terms of overlaps of  light-front wave functions (LFWFs), which were calculated analytically in light-front Hamiltonian perturbation theory. In \cite{Mukherjee14} we calculated the quark Wigner distributions for the unpolarized and longitudinally polarized target. The results were found to depend on a regulator on the momentum transfer by the target in the transverse direction. The convergence was improved, and the regulator dependence was removed in  the quark Wigner distributions in \cite{More17} and transverse  polarization was included. Here we continue our study and calculate the gluon Wigner distributions, using improved numerical convergence, and remove the regulator dependence present in the earlier work \cite{Mukherjee15}. We also include transverse and linear polarization in the study of gluon Wigner distributions at leading twist.                
 
The manuscript is organized in the following manner. In Sec. \ref{WD}, we begin with the field theory definition of the gluon Wigner distribution. 
We give the analytical expressions for gluon Wigner distribution at leading twist in the dressed quark model.
In Sec. \ref{numerical}, we explain the numerical strategy used for studying the 
Wigner distribution followed by the discussion of our numerical results. In Sec. \ref{density} We discuss spin density in transverse momentum and transverse position space. We present a discussion of the small-$x$ behaviour of the quark and  gluon helicities and orbital angular momentum in our model in Sec. \ref{smallx}. 
Conclusions are given in Sec. \ref{conclusion}. 

\section{Gluon Wigner Distributions in dressed quark model}\label{WD}
The Wigner distribution of gluon can be defined as 
\cite{Meissner09, Lorce11} 
\ba
x W_{\sigma, \sigma'}(x,{\bs k_\perp},{\bs b_\perp})\es\int \frac{d^2 {\bs \Delta_\perp}}{(2\pi)^2}e^{-i{\bs \Delta_\perp}.{\bs b_\perp}} \int 
\frac{dz^{-}d^{2} {\bs z}_{\perp}}{2(2\pi)^3 p^+}e^{i k.z} \nn 
 &\times&\Big{\langle } p^{+},-\frac{{\bs \Delta_\perp}}{2},\sigma' \Big{|}
\Gamma^{ij} F^{+i}\Big( -\frac{z}{2}\Big) F^{+j}\Big( \frac{z}{2}\Big) \Big{|}
p^{+},\frac{{\bs \Delta_\perp}}{2},\sigma  \Big{\rangle }  \Big{|}_{z^{+}=0}
\label{wignerdistribution} 
\ea
where  ${\bs \Delta_\perp}$ and ${\bs b_\perp}$ is the transverse momentum transfer from the target state and 
 impact parameter space variable respectively. ${\bs b_\perp}$ is conjugate to ${\bs \Delta_\perp}$,
the gluon  field strength tensor is given by 
\ba
F_a^{+i} = \partial^+ A_a^i -  \partial^i  A_a^+ + g f_{abc} A_b^+ A_c^i
\ea
The operator structure for gluon at twist two are \cite{Lorce13}
(i) $\Gamma^{ij}=\delta_\perp^{ij}$,~~
(ii) $\Gamma^{ij} =-i\epsilon^{ij}_{\perp}$ ~~~
(iii) $\Gamma^{ij}=\Gamma^{RR}$ and 
(iv) $\Gamma^{ij}=\Gamma^{LL}$, where $L(R)$ are left(right) polarization of the gluon (to be defined later). 
We have suppressed the color indices. 
The above  correlator needs two gauge links for color gauge invariance. We choose light-cone gauge and take the gauge link to be unity, that is we will not be calculating the effect of the transverse link at light-cone infinity.   
In the previous work \cite{More17}, we discussed
the quark Wigner distribution for the dressed quark model. The dressed quark state with momentum `$p$' and helicity `$s$' can be written in terms of light-front wave functions (LFWFs) as the expansion of the  state in Fock space: 
\ba
  \Big{| }p^{+}, {\bs p}_{\perp}, s \Big{\rangle} &=& \Phi^{s}(p) b^{\dagger}_{s}(p) | 0 \rangle +
 \sum_{s_1 s_2} \int \frac{dp_1^{+}d^{2}{\bs p}_1^{\perp}}{ \sqrt{16 \pi^3 p_1^{+}}}
 \int \frac{dp_2^{+}d^{2}{\bs p}_2^{\perp}}{ \sqrt{16 \pi^3 p_2^{+}}} \sqrt{16 \pi^3 p^{+}}\,\,
\nn 
 &\times& \delta^3(p-p_1-p_2) \Phi^{s}_{s_1 s_2}(p;p_1,p_2) b^{\dagger}_{s_1}(p_1) 
 a^{\dagger}_{s_2}(p_2)  | 0 \rangle 
 \ea
where $\Phi^{s}_{s_1 s_2}(p;p_1,p_2)$ is the two-particle  LFWF. $\Phi^{s}(p)$ gives the wave function normalization \cite{Hari99}. $\Phi^{s}_{s_1 s_2}(p;p_1,p_2)$ 
gives the probability amplitude to find a bare quark (gluon) with momentum $p_1 (p_2)$ and helicity $s_1 (s_2)$ inside the dressed quark. Using the Jacobi momenta
\ba k_{i}^{+} = x_{i}P^+ ~\text{and}~~~ {\bs k}_{i}^{\perp} ={\bs q}_{i}^{\perp} +  x_{i}{\bs P}^{\perp} \ea
so that 
\ba\sum_i x_i=1,~~~~~\sum_i {\bs q}_{i\perp}=0\ea 
the two-particle LFWF can be written in terms of the boost-invariant variables  as
\ba\sqrt{P^+}\Phi(p; p_1, p_2) = \Psi(x_{i},{\bs q}_{i}^{\perp})\ea
${\bs k}_\perp$ is the average transverse momentum of the gluon and $x$ is the momentum fraction of the gluon. 
This LFWF can be calculated analytically using light-front Hamiltonian perturbation theory \cite{Hari99}. 
The gluon-gluon correlators in terms of overlap of two-particle LFWFs are given by
\begin{itemize}
 \item[i)]  For $\Gamma^{ij} = \delta_\perp ^{ij}$
\ba\label{W1}
\mathcal{W}^{1}_{\sigma \sigma'}(x,{\bs k _\perp},{\bs \Delta _\perp})=-\!\!\sum_{\sigma_1, \lambda_{1},\lambda_{2}}
\Bigg[\Psi^{*\sigma'}_{\sigma_1 \lambda_{1} }(\hat{x},\hat{{\bs q}}'_{\perp}) \Psi^{\sigma}_{\sigma_1 \lambda_2}(\hat{x},\hat{{\bs q}}_{\perp})
\Big(\epsilon_{\lambda_2}^{1} \epsilon_{\lambda_1}^{*1}+\epsilon_{\lambda_2}^{2} \epsilon_{\lambda_1}^{*2}\Big)\Bigg]
\ea
\item[ii)] For $\Gamma^{ij} = -i\epsilon^{ij}_{\perp}$ 
\ba\label{W2}
\mathcal{W}^{2}_{\sigma \sigma'}(x,{\bs k _\perp},{\bs \Delta _\perp}) = -i\!\!\!\sum_{\sigma_1, \lambda_{1},\lambda_{2}}
\Bigg[ \Psi^{*\sigma'}_{\sigma_1 \lambda_{1} }(\hat{x},\hat{\bs q}'_{\perp}) \Psi^{\sigma}_{\sigma_1 \lambda_2}(\hat{x},\hat{{\bs q}}_{\perp})
\Big(\epsilon_{\lambda_2}^{1} \epsilon_{\lambda_1}^{*2}-\epsilon_{\lambda_2}^{2} \epsilon_{\lambda_1}^{*1}\Big)\Bigg]
\ea
\item[iii)] For $\Gamma^{RR} $
\ba\label{W3}
\mathcal{W}^{3}_{\sigma \sigma'}(x,{\bs k _\perp},{\bs \Delta _\perp}) = -\!\!\!\sum_{\sigma_1, \lambda_{1},\lambda_{2}}
\Bigg[ \Psi^{*\sigma'}_{\sigma_1 \lambda_{1} }(\hat{x},\hat{{\bs q}}'_{\perp}) \Psi^{\sigma}_{\sigma_1 \lambda_2}(\hat{x},\hat{{\bs q}}_{\perp})
\epsilon_{\lambda_2}^{R} \epsilon_{\lambda_1}^{*R}\Bigg]
\ea
\item[iv)] For $\Gamma^{LL} $
\ba\label{W4}
\mathcal{W}^{4}_{\sigma \sigma'}(x,{\bs k _\perp},{\bs \Delta _\perp}) = -\!\!\!\sum_{\sigma_1, \lambda_{1},\lambda_{2}}
\Bigg[ \Psi^{*\sigma'}_{\sigma_1 \lambda_{1} }(\hat{x},\hat{{\bs q}}'_{\perp}) \Psi^{\sigma}_{\sigma_1 \lambda_2}(\hat{x},\hat{{\bs q}}_{\perp})
\epsilon_{\lambda_2}^{L} \epsilon_{\lambda_1}^{*L}\Bigg]
\ea
\end{itemize}
where $ \hat{x} = (1-x)$,~ $\hat{{\bs q}}_{\perp}=-{\bs q}_{\perp}$, \, ${\bs q}_\perp= {\bs k}_\perp+{{\bs \Delta}_\perp\over 2} (1-x), \,\, {\bs q}'_\perp= {\bs k}_\perp-{{\bs \Delta}_\perp\over 2} (1-x).$ We choose the gluon right (left) polarization as~$\epsilon_{\lambda}^{R(L)}=\epsilon_{\lambda}^1 \pm i\epsilon_{\lambda}^2$. 
Wigner distribution is obtained  by taking the Fourier transform of Eqs.~(\ref{W1})-(\ref{W4}) and is given by 
\ba
x W^{\alpha} (x,{\bs k _\perp},{\bs b _\perp}, \hat{\bs e})=\int\frac{d^2 {\bs \Delta _\perp}}{2(2\pi)^2}\,\,e^{-i{\bs \Delta _\perp}\cdot {\bs b _\perp}}\,\,
{\mathcal{W}}^{\alpha}_{\sigma \sigma}(x,{\bs k _\perp},{\bs \Delta _\perp}) 
\ea
with $\alpha=1,2,3,4$. 
Thus, for different polarization
of gluon viz (U) unpolarized, (L) longitudinally polarized and two ($\T$) linearly polarized, one obtains 16 gluon Wigner distributions at leading twist. In the above expression, the polarization of the target state is denoted by $\hat{\bs e}$.  $T$ denotes the transverse polarization of the target. 
We denote the gluon Wigner distribution as $W_{\lambda\lambda^\prime}$,
where $\lambda$ and $\lambda^\prime$ are the polarization of target state and gluon respectively. We categorize 16 Wigner distributions for different polarization combinations at leading twist.

\subsection{Unpolarized target and different gluon polarization}
%
\noindent
The unpolarized Wigner distribution 
\ba
W_{UU}(x, {\bs k _\perp}, {\bs b _\perp})&=&\frac1{2}\Big[W^1(x, {\bs k _\perp}, {\bs b _\perp},\hat{\bs e}_z)+W^1(x, {\bs k _\perp}, {\bs b _\perp},-\hat{\bs e}_z)\Big]
\label{rhouu} 
\ea
The unpolarized-longitudinally polarized Wigner distribution
\ba
W_{UL}(x, {\bs k _\perp}, {\bs b _\perp})&=&\frac1{2}\Big[W^2(x, {\bs k _\perp}, {\bs b _\perp},\hat{\bs e}_z)+W^2(x, {\bs k _\perp}, {\bs b _\perp},-\hat{\bs e}_z)\Big]
\label{rhoul}
\ea
The unpolarized-linearly polarized Wigner distribution
\ba
W_{U \T}^{(R)}(x, {\bs k _\perp}, {\bs b _\perp})&=&\frac1{2}\Big[W^3(x, {\bs k _\perp}, {\bs b _\perp},\hat{\bs e}_z)+W^3(x, {\bs k _\perp}, {\bs b _\perp},-\hat{\bs e}_z)\Big]
\ea
\ba\label{wutL}
W_{U \T}^{(L)}(x, {\bs k _\perp}, {\bs b _\perp})&=&\frac1{2}\Big[W^4(x, {\bs k _\perp}, {\bs b _\perp},\hat{\bs e}_z)+W^4(x, {\bs k _\perp}, {\bs b _\perp},-\hat{\bs e}_z)\Big]
\ea
The superscipt `$L(R)$' represents left (right) polarization of gluon.
\subsection{Longitudinal polarized target and different gluon polarization}
\noindent
The longitudinal-unpolarized Wigner distribution
\ba
W_{LU}(x, {\bs k _\perp}, {\bs b _\perp})&=&\frac1{2}\Big[W^1(x, {\bs k _\perp}, {\bs b _\perp},\hat{\bs e}_z)-W^1(x, {\bs k _\perp}, {\bs b _\perp},-\hat{\bs e}_z)\Big]
\label{rholu}
\ea
The longitudinal Wigner distribution
\ba
W_{LL}(x, {\bs k _\perp}, {\bs b _\perp})&=&\frac1{2}\Big[W^2(x, {\bs k _\perp}, {\bs b _\perp},\hat{\bs e}_z)-W^2(x, {\bs k _\perp}, {\bs b _\perp},-\hat{\bs e}_z)\Big]
\label{rholl}
\ea
The longitudinal-linearly polarized Wigner distribution
\ba
W^{(R)}_{L\T}(x, {\bs k _\perp}, {\bs b _\perp})&=&\frac1{2}\Big[W^{3}(x, {\bs k _\perp}, {\bs b _\perp},\hat{\bs e}_z)-W^{3}(x, {\bs k _\perp}, {\bs b _\perp},-\hat{\bs e}_z)\Big]
\ea
\ba
W^{(L)}_{L\T}(x, {\bs k _\perp}, {\bs b _\perp})&=&\frac1{2}\Big[W^{4}(x, {\bs k _\perp}, {\bs b _\perp},\hat{\bs e}_z)-W^{4}(x, {\bs k _\perp}, {\bs b _\perp},-\hat{\bs e}_z)\Big]
\ea
\subsection{Transversely polarized target and different gluon polarization}
\noindent
The transversely polarized unpolarized  Wigner distribution
\ba
W^i_{T U}(x, {\bs k _\perp}, {\bs b _\perp})&=&\frac1{2}\Big[W^1(x, {\bs k _\perp}, {\bs b _\perp},\hat{\bs e}_i)-W^1(x, {\bs k _\perp}, {\bs b _\perp},-\hat{\bs e}_i)\Big]
\ea
The transversely-longitudinally polarized  Wigner distribution
\ba
W^i_{T L}(x, {\bs k _\perp}, {\bs b _\perp})&=&\frac1{2}\Big[W^2(x, {\bs k _\perp}, {\bs b _\perp},\hat{\bs e}_i)-W^2(x, {\bs k _\perp}, {\bs b _\perp},-\hat{\bs e}_i)\Big]
\ea
The transversely-linearly polarized  Wigner distribution
\ba
W^{i(R)}_{T \T}(x, {\bs k _\perp}, {\bs b _\perp})&=&\frac1{2}\Big[W^{3}(x, {\bs k _\perp}, {\bs b _\perp},\hat{\bs e}_i)-W^{3}(x, {\bs k _\perp}, {\bs b _\perp},-\hat{\bs e}_i)\Big]
\label{rhott}
\ea
\ba
W^{i(L)}_{T \T}(x, {\bs k _\perp}, {\bs b _\perp})&=&\frac1{2}\Big[W^{4}(x, {\bs k _\perp}, {\bs b _\perp},\hat{\bs e}_i)-W^{4}(x, {\bs k _\perp}, {\bs b _\perp},-\hat{\bs e}_i)\Big]
\label{rhott}
\ea
where  $i$ denotes the transverse directions. $\hat{\bs e}_i$ correspond to the transverse polarization of the target state and these can be expressed as a linear combination of helicity states. We have chosen the transverse polarization to be in $x$ direction.

Using the analytic expression of the two-particle LFWFs, we calculate the Wigner distributions. Expressions for the unpolarized and longitudinal  polarization have been given earlier in \cite{Mukherjee15}.  For completeness, here we give them again and include the expressions for transverse/linear polarizations also. 
The analytical expression for six linearly independent gluon Wigner distributions are given as:
\ba
W_{UU}(x, {\bs k _\perp}, {\bs b _\perp})&=&N\int \frac{d^2 {\bs \Delta _\perp}}{2(2\pi)^2} \frac{\cos({\bs \Delta _\perp}\cdot{\bs b _\perp})}{D({\bs q_\perp})D({\bs q'_\perp})}\nn
&\times&\left[-\frac{4 m^2 x^4+(x^2-2x+2) \Big(4{\bs k _\perp}^2-{\bs \Delta _\perp}^2 (1-x)^2\Big)}{(1-x)^2 x^3}\right]\label{uu}\\[2ex]
W_{UL}(x, {\bs k _\perp}, {\bs b _\perp})&=&N \int \frac{d^2 {\bs \Delta _\perp}}{2(2\pi)^2} \frac{\sin({\bs \Delta _\perp}\cdot{\bs b _\perp})}{D({\bs q_\perp})D({\bs q'_\perp})}
\frac{4 \left(x^2-2 x+2\right) (\Delta_y k_x-\Delta_x k_y)}{(1-x) x^3}\label{ul}\nn \\[2ex]
W_{LU}(x, {\bs k _\perp}, {\bs b _\perp})&=&N \int \frac{d^2 {\bs \Delta _\perp}}{2(2\pi)^2} \frac{\sin({\bs \Delta _\perp}\cdot{\bs b _\perp})}{D({\bs q_\perp})D({\bs q'_\perp})}\,\,\,
\frac{4 (2-x) (\Delta_y k_x-\Delta_x k_y)}{(1-x) x^2}\label{lu}\\[2ex]
W_{LL}(x, {\bs k _\perp}, {\bs b _\perp})&=&N \int \frac{d^2 {\bs \Delta _\perp}}{2(2\pi)^2} \frac{\cos({\bs \Delta _\perp}\cdot{\bs b _\perp})}{D({\bs q_\perp})D({\bs q'_\perp})}\nn
&\times&\left[-\frac{4 m^2 x^3+(2-x) \Big(4{\bs k _\perp}^2-{\bs \Delta _\perp}^2 (1-x)^2\Big)}
{(1-x)^2 x^2}\right]
\label{ll}\\
W^x_{T U}(x, {\bs k _\perp}, {\bs b _\perp})&=&N \int \frac{d^2 {\bs \Delta _\perp}}{2(2\pi)^2} \frac{\sin({\bs \Delta _\perp\cdot}{\bs b _\perp})}{D({\bs q_\perp})D({\bs q'_\perp})}\,\,\,
\frac{4 m \Delta_x}{x}\label{tu}\\[2ex]
W^x_{T L}(x, {\bs k _\perp}, {\bs b _\perp})&=&N \int \frac{d^2 {\bs \Delta _\perp}}{2(2\pi)^2} \frac{\cos({\bs \Delta _\perp}\cdot{\bs b _\perp})}{D({\bs q_\perp})D({\bs q'_\perp})}\,\,\,
\frac{8 m k_y}
{x (1-x)}\label{tl}
\ea
where,
$\,\,N=\frac{g^2 C_F}{2 (2 \pi)^2},~~C_F \text{ is the color factor}$
\ba 
D({\bs q_\perp})=\left[m^2 - \frac{m^2 + ({ {\bs k _\perp}}+\frac{{ {\bs \Delta _\perp}} (1-x)}{2})^2 }{1-x} - \frac{({ {\bs k _\perp}}+\frac{{ {\bs \Delta _\perp}} (1-x)}{2})^2}{x} \right]\nn\\
D({\bs q'_\perp})=\left[m^2 - \frac{m^2 + ({ {\bs k _\perp}}-\frac{{ {\bs \Delta _\perp}} (1-x)}{2})^2 }{1-x} - \frac{({ {\bs k _\perp}}-\frac{{ {\bs \Delta _\perp}} (1-x)}{2})^2}{x} \right]
\ea
The remaining six gluon Wigner distributions can be expressed as linear combinations of the above six distributions in this model as:
\ba
W^{(R)}_{U \T}(x, {\bs k _\perp}, {\bs b _\perp})
&=&W_{UU}(x, {\bs k _\perp}, {\bs b _\perp})-W_{UL}(x, {\bs k _\perp}, {\bs b _\perp})\label{wutr}\\[2ex]
W^{(L)}_{U \T}(x, {\bs k _\perp}, {\bs b _\perp})
&=&W_{UU}(x, {\bs k _\perp}, {\bs b _\perp})+W_{UL}(x, {\bs k _\perp}, {\bs b _\perp})\label{wutl}\\[2ex]
W^{(R)}_{L \T}(x, {\bs k _\perp}, {\bs b _\perp})
&=& -W_{LL}(x, {\bs k _\perp}, {\bs b _\perp})+W_{LU}(x, {\bs k _\perp}, {\bs b _\perp})\label{wltr}\\[2ex]
W^{(L)}_{L \T}(x, {\bs k _\perp}, {\bs b _\perp})
&=&W_{LL}(x, {\bs k _\perp}, {\bs b _\perp})+W_{LU}(x, {\bs k _\perp}, {\bs b _\perp})\label{wltl}\\[2ex]
W^{(R)x}_{T \T}(x, {\bs k _\perp}, {\bs b _\perp})
&=&-W^x_{T L}(x, {\bs k _\perp}, {\bs b _\perp})+W^x_{T U}(x, {\bs k _\perp}, {\bs b _\perp})\label{wttr}\\[2ex]
W^{(L)x}_{T \T}(x, {\bs k _\perp}, {\bs b _\perp})
&=&W^x_{T L}(x, {\bs k _\perp}, {\bs b _\perp})+W^x_{T U}(x, {\bs k _\perp}, {\bs b _\perp})\label{wttl}
\ea
\section{Numerical calculations}\label{numerical}
\begin{figure}[h] 
(a)\includegraphics[width=7.5cm,height=4.8cm]{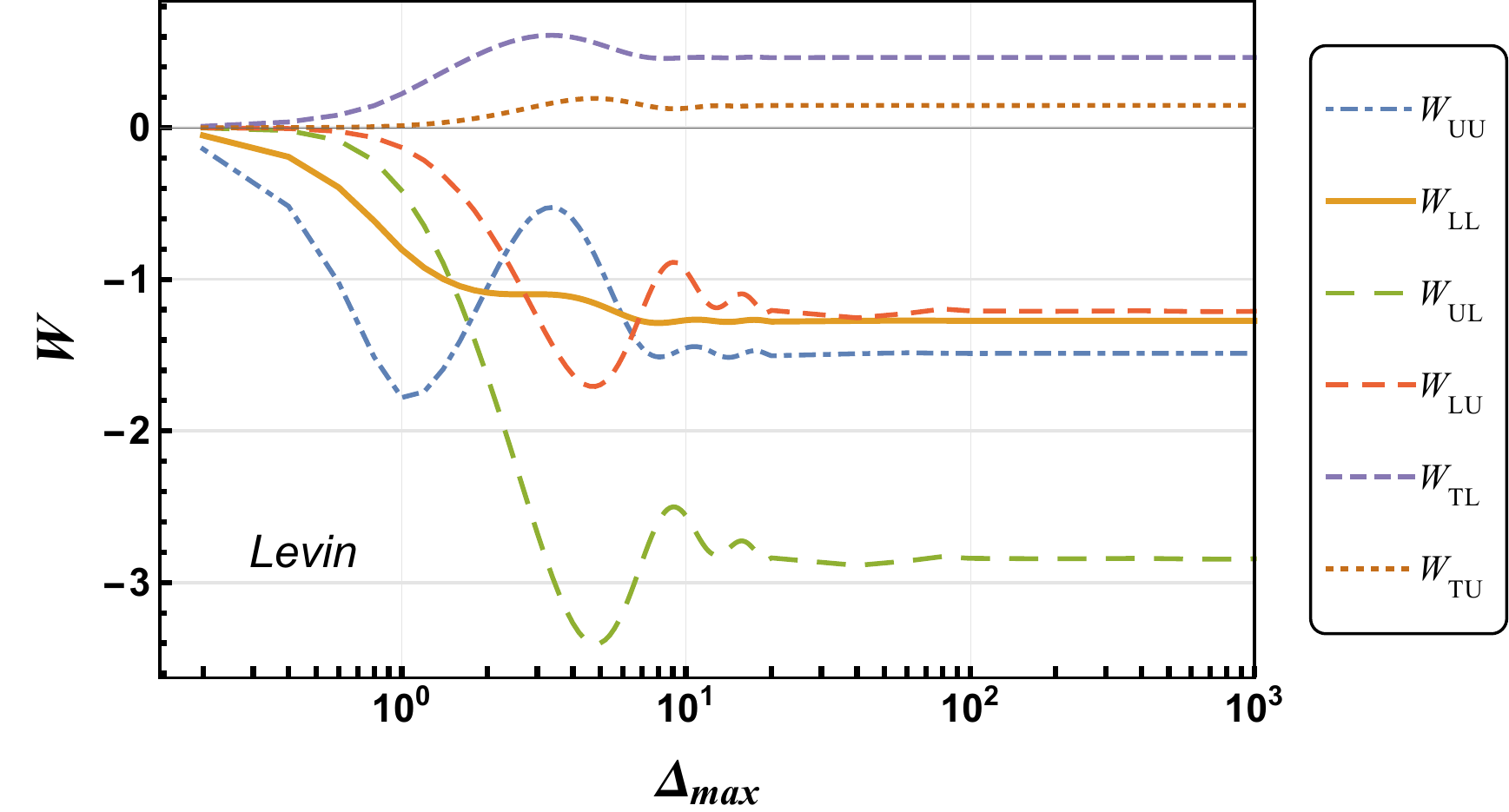}(b)\includegraphics[width=7.5cm,height=4.8cm]{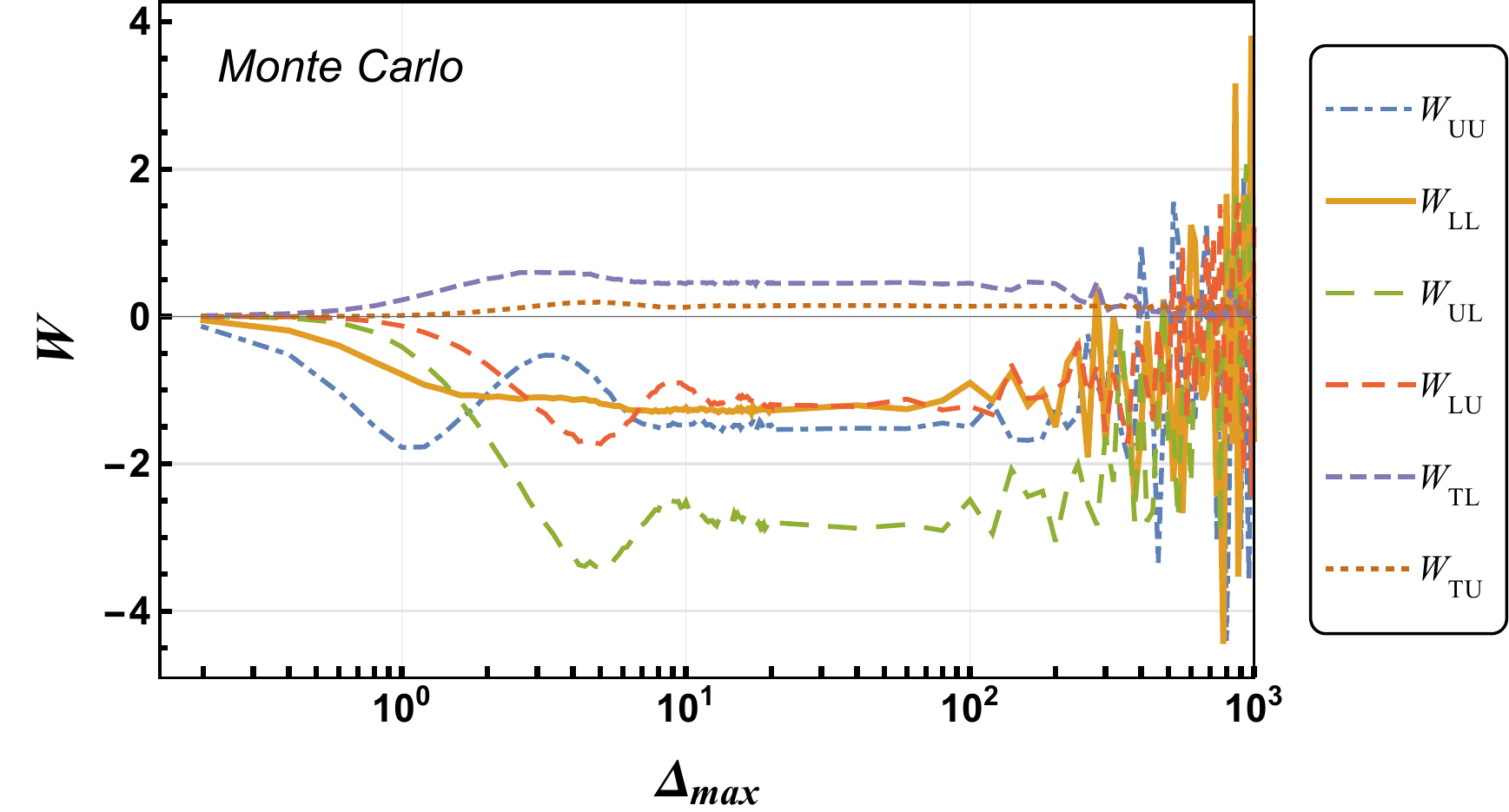}
  \caption{Plot of six Wigner distributions vs $\Delta_{max}(\mathrm{GeV})$ at a fixed value of
  $b_x=0.4~\mathrm{GeV}^{-1} , b_y=0.5~\mathrm{GeV}^{-1} , k_x=0.0~\mathrm{GeV}$ and
  $k_y=0.4~\mathrm{GeV}$ using the Levin (a) and Monte Carlo (b) integration methods.}  \label{convPlot}
\end{figure}

The gluon Wigner distribution is a five-dimensional phase space distribution with two 
transverse components (${\bs k}_\perp, {\bs b}_\perp$) and a longitudinal momentum fraction  ($x$). In the following, we integrate out the longitudinal momentum fraction ($x$)
and study its behavior in the transverse phase space (${\bs k _\perp}, {\bs b _\perp}$).
Figure.~\ref{convPlot} shows the plot for the six linearly independent gluon Wigner distributions as a function of $\Delta_{max}$, which is the upper cutoff of the $\Delta_\perp$ integration. Figure.~\ref{convPlot}~(a) shows the results when we apply Levin method \cite{Levin82, Levin96, Levin97}, whereas Fig.~\ref{convPlot}~(b) is generated using Monte Carlo (MC) method. Both plots are for the fixed value of 
 $b_x=0.4~\mathrm{GeV}^{-1} , b_y=0.5~\mathrm{GeV}^{-1} , k_x=0.0~\mathrm{GeV}$ and
  $k_y=0.4~\mathrm{GeV}$ with $\Delta_{max}$ going from 1 upto $10^3~\mathrm{GeV}$. For all plots we use $m=0.33~\mathrm{GeV}$. It is clear that for $\Delta_{max}=20~ \mathrm{GeV}$ and beyond the results are independent of this cutoff when Levin method is used, whereas in the MC approach they become highly oscillatory for higher values of the cutoff.  This imposed a serious restriction in our previous work \cite{Mukherjee15} where MC method was used. In this paper we have used Levin method and $\Delta_{max}=20~\mathrm{GeV}$, as we did in \cite{More17}.

\begin{figure}[h!] 
(a)\includegraphics[width=7.5cm,height=5.5cm]{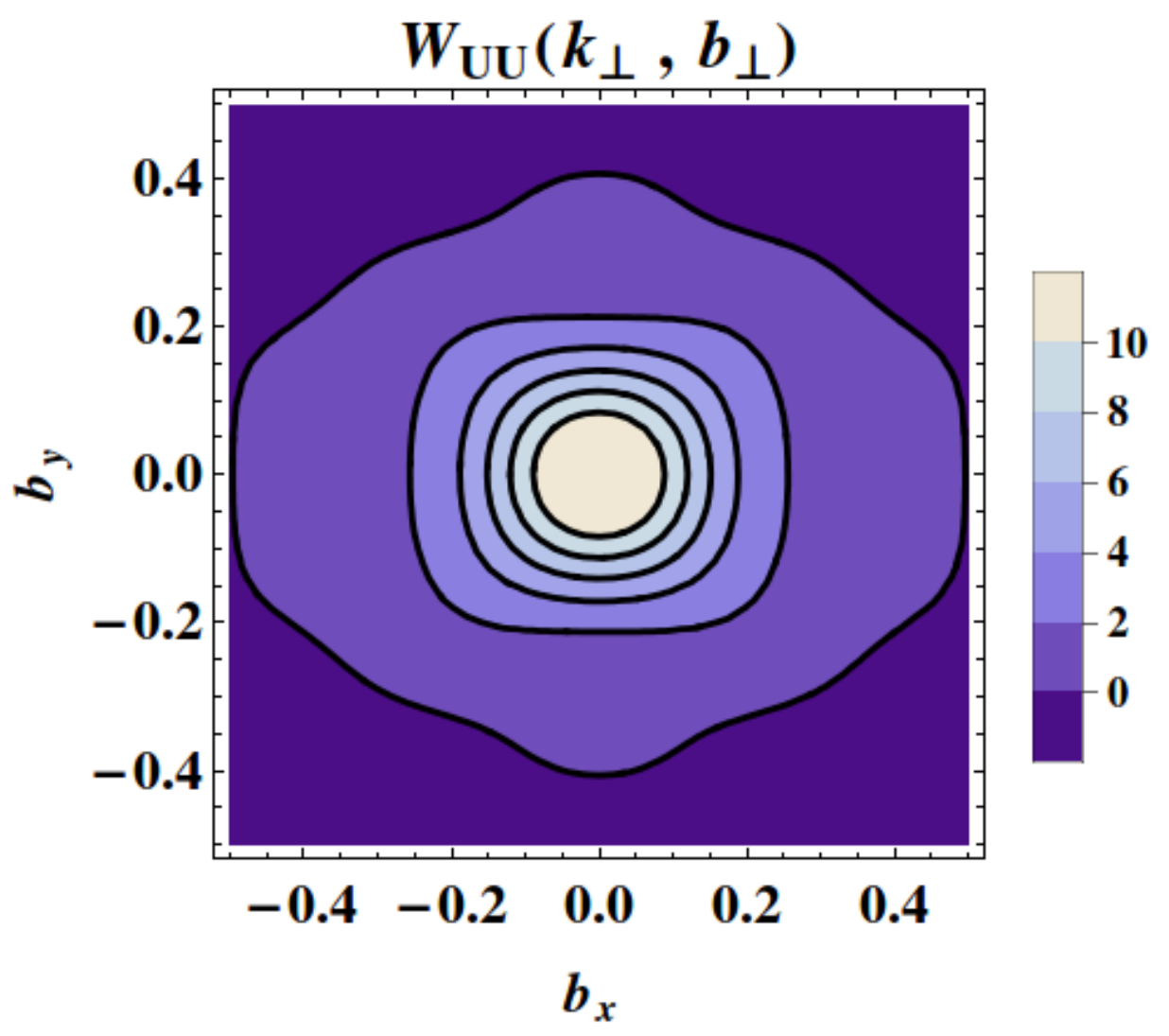}(d)\includegraphics[width=7.5cm,height=5.5cm]{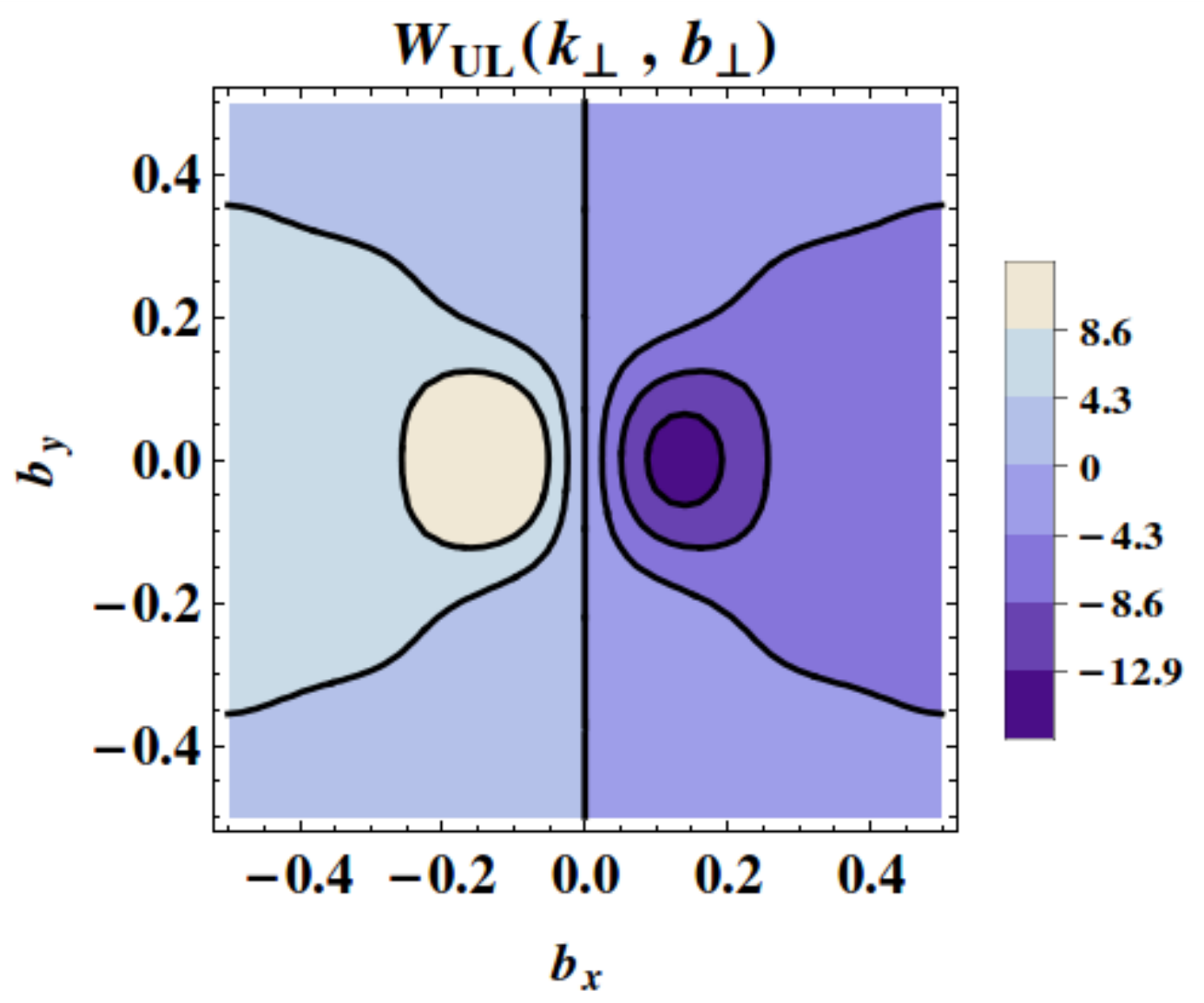}\\[5ex]
(b)\includegraphics[width=7.5cm,height=5.5cm]{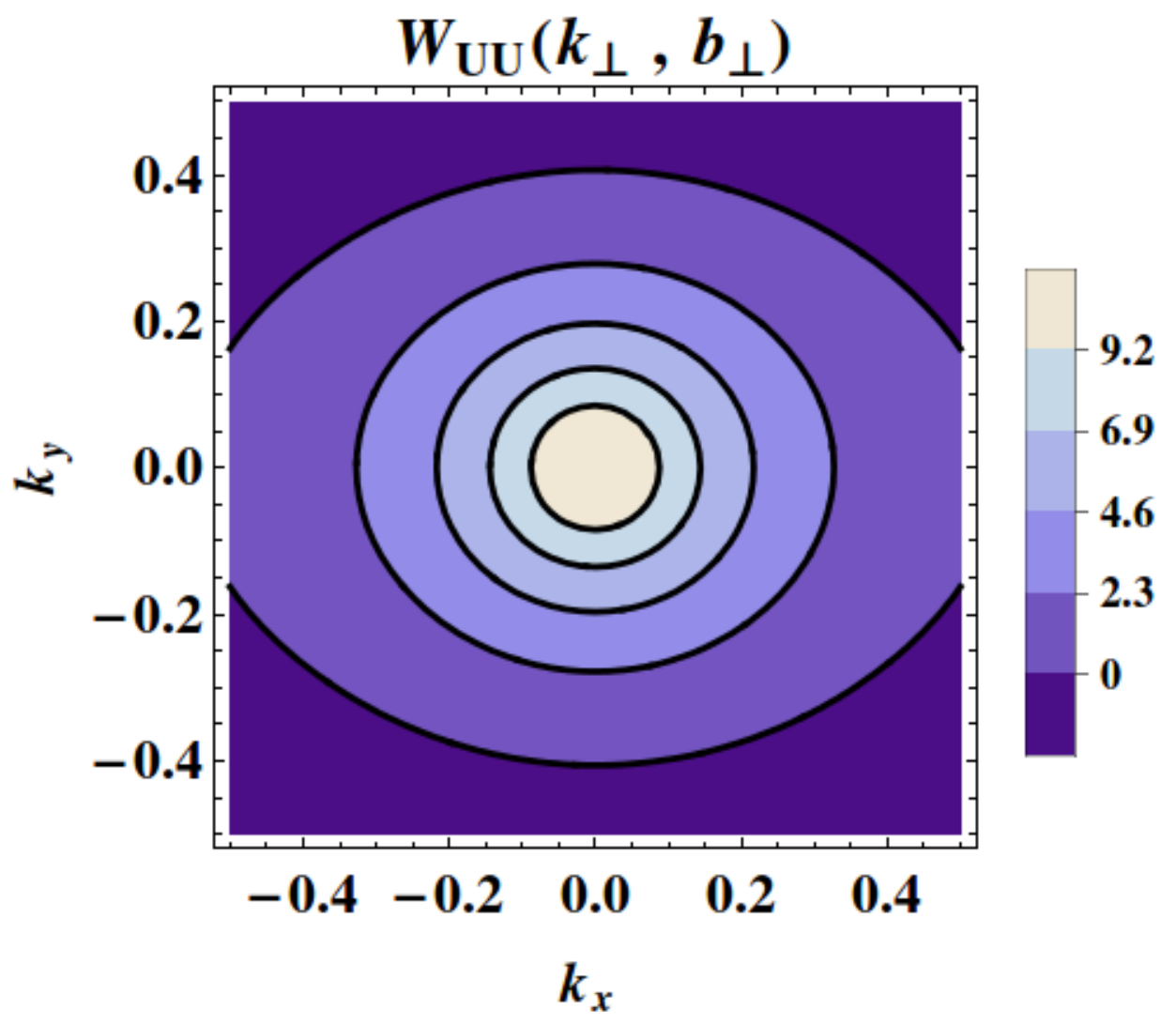}(e)\includegraphics[width=7.5cm,height=5.5cm]{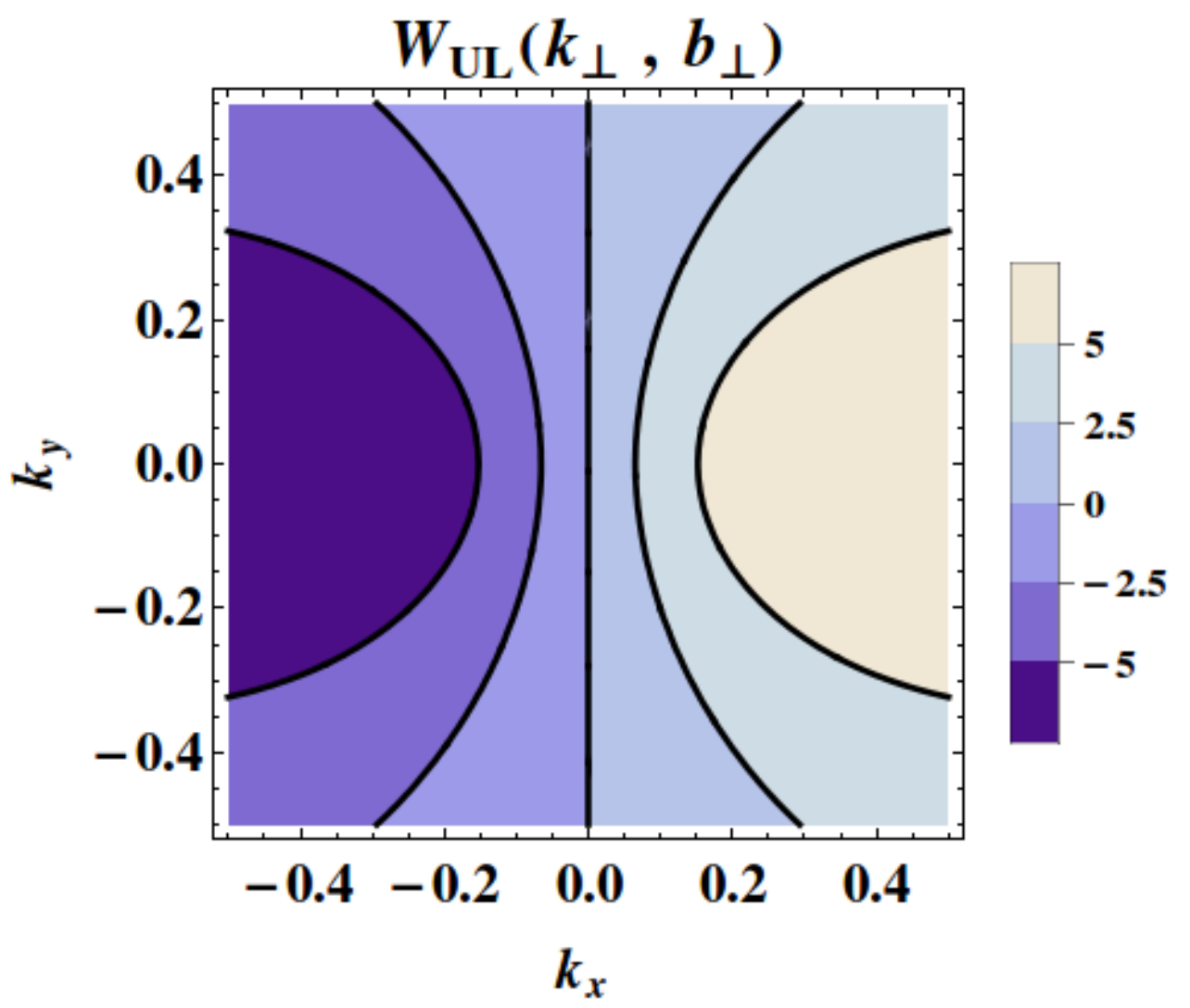}\\[5ex]
(c)\includegraphics[width=7.5cm,height=5.5cm]{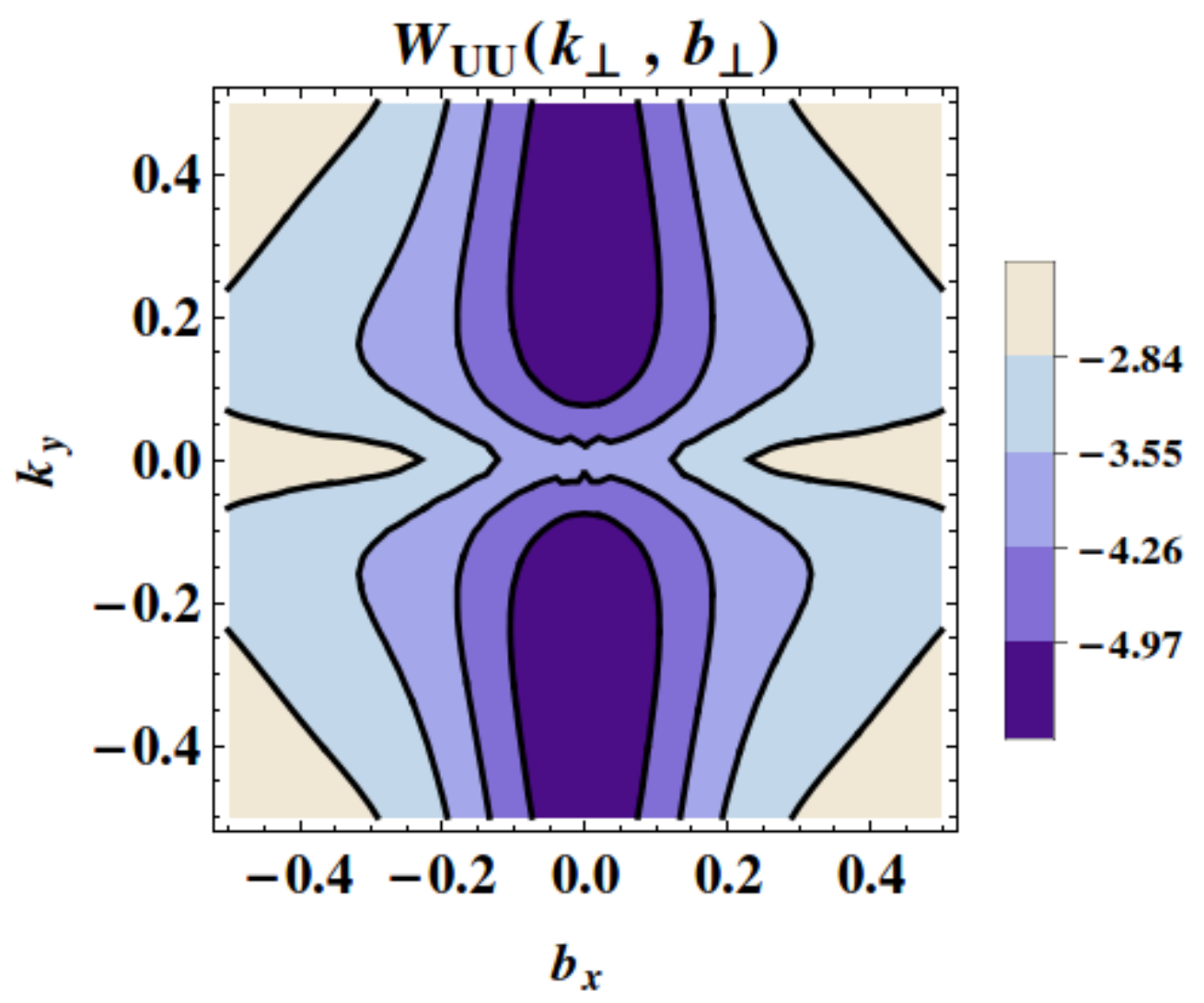}(f)\includegraphics[width=7.5cm,height=5.5cm]{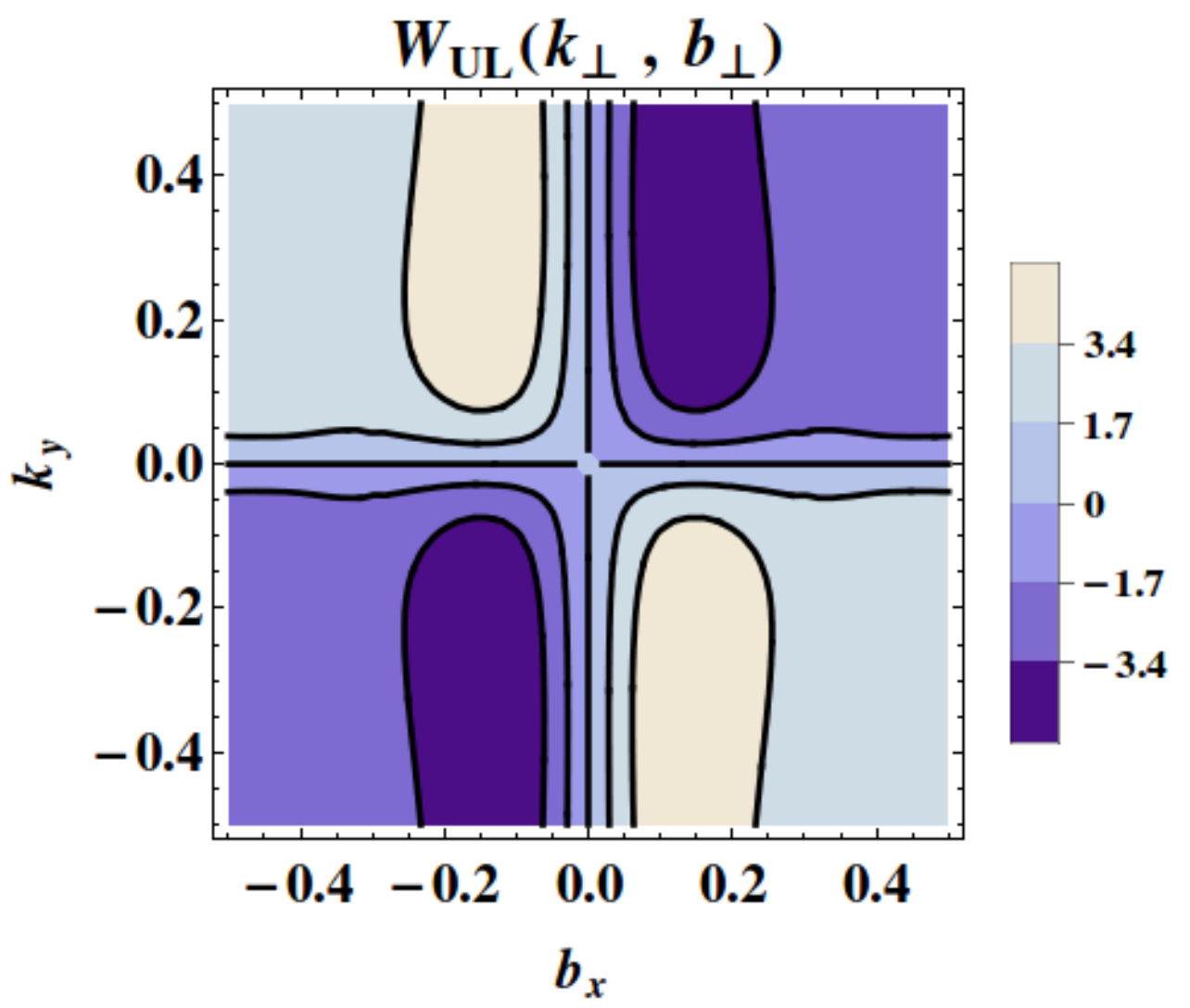}  
  \caption{ Plot of gluon Wigner distributions $W_{UU}({\bs k}_\perp, {\bs b}_\perp)$ and 
  $W_{UL}({\bs k}_\perp, {\bs b}_\perp)$ at $\Delta_{max} = 20~\mathrm{GeV}$. The first row displays the 
  two distributions in ${\bs b}_\perp$-space with ${\bs k}_\perp= 0.4~\mathrm{GeV} \,\hat{{\bs e}}_y$. The second row shows the two distributions in ${\bs k}_\perp$-space with ${\bs b}_\perp= 0.4~\mathrm{GeV}^{-1}\, \hat{{\bs e}}_y$.  The last row represents the two distributions in mixed space.}  \label{wuu}
\end{figure}

\begin{figure}[h!]  
(a)\includegraphics[width=7.5cm,height=5.5cm]{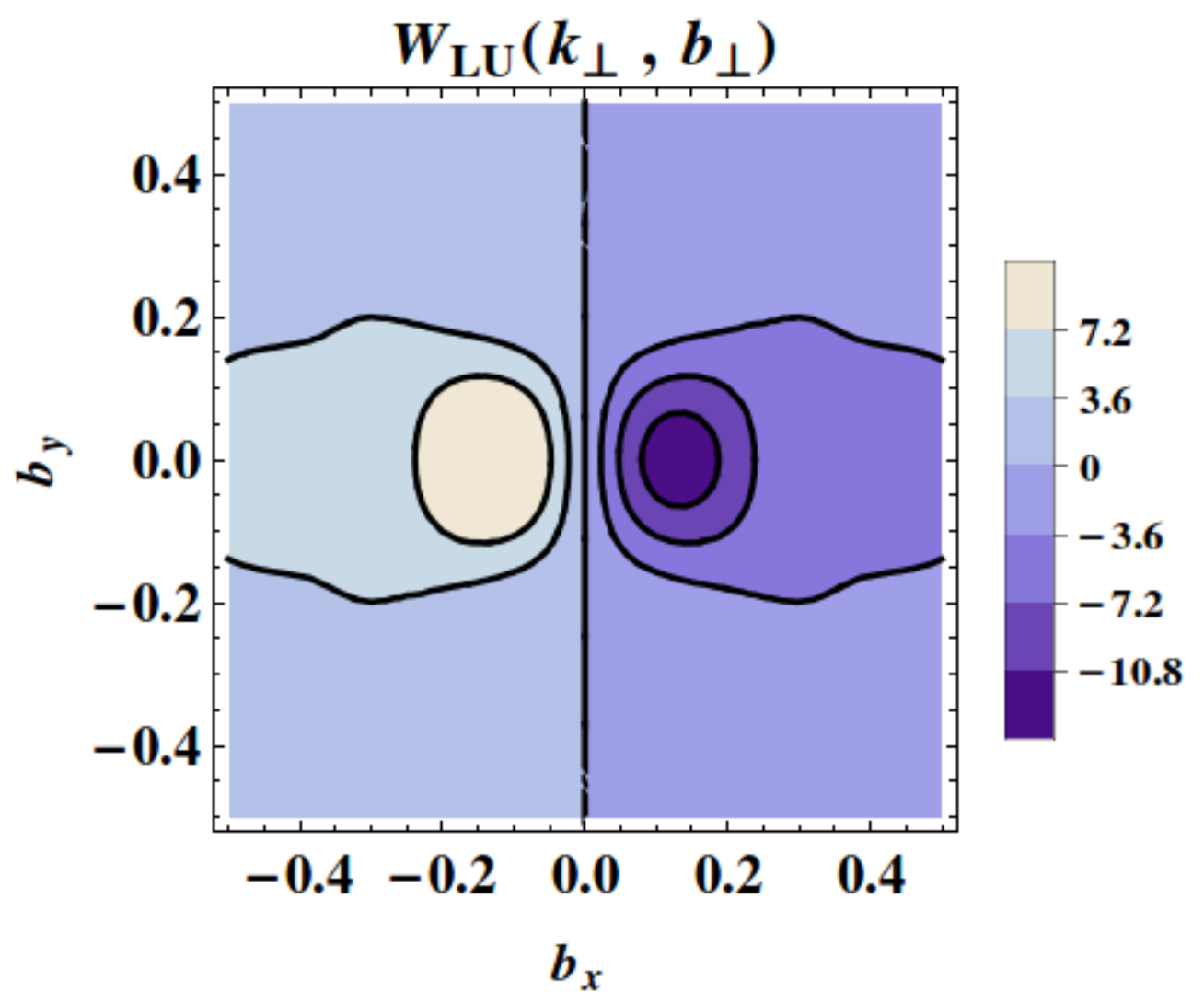}(d)\includegraphics[width=7.5cm,height=5.5cm]{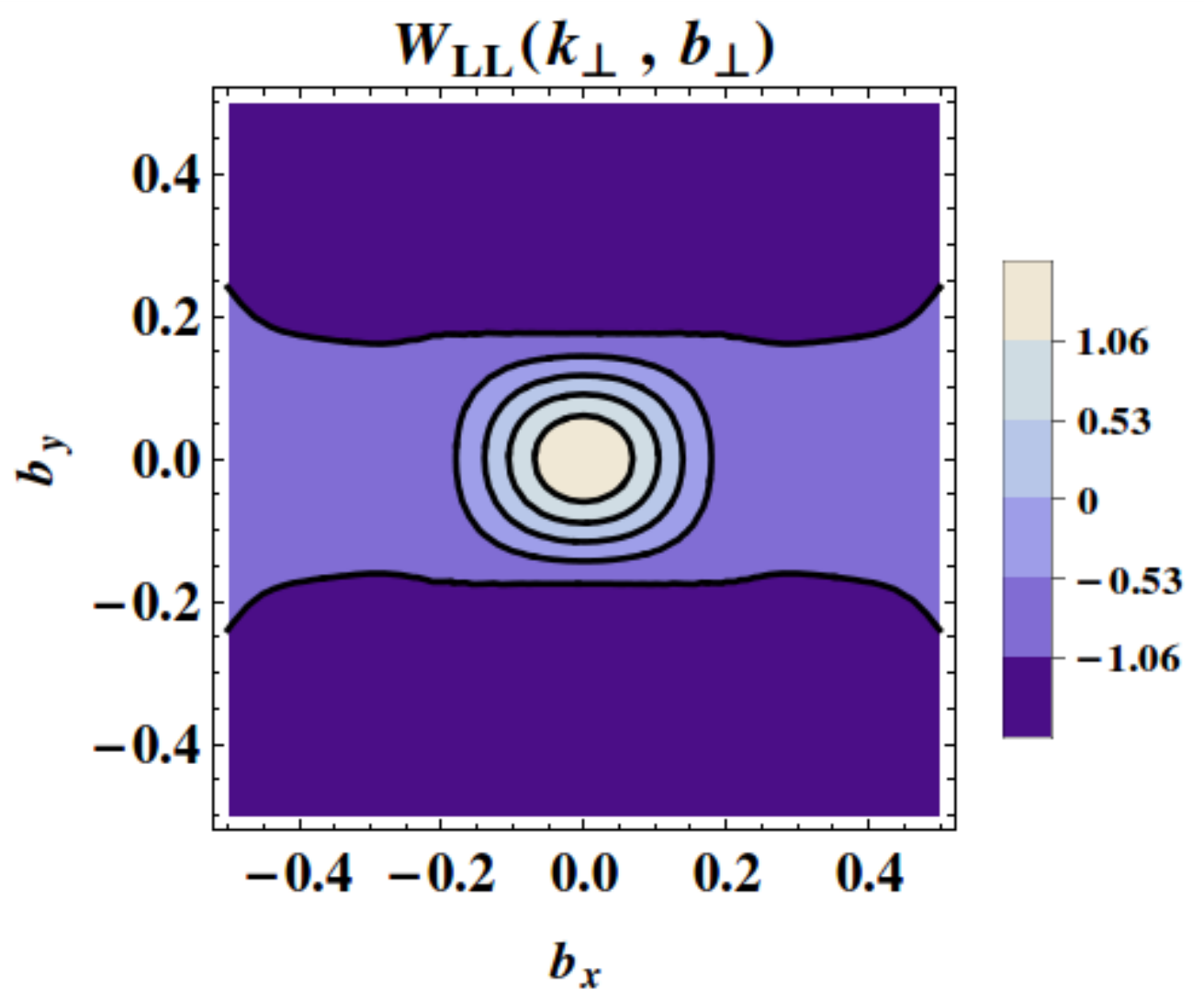}\\[2ex]
(b)\includegraphics[width=7.5cm,height=5.5cm]{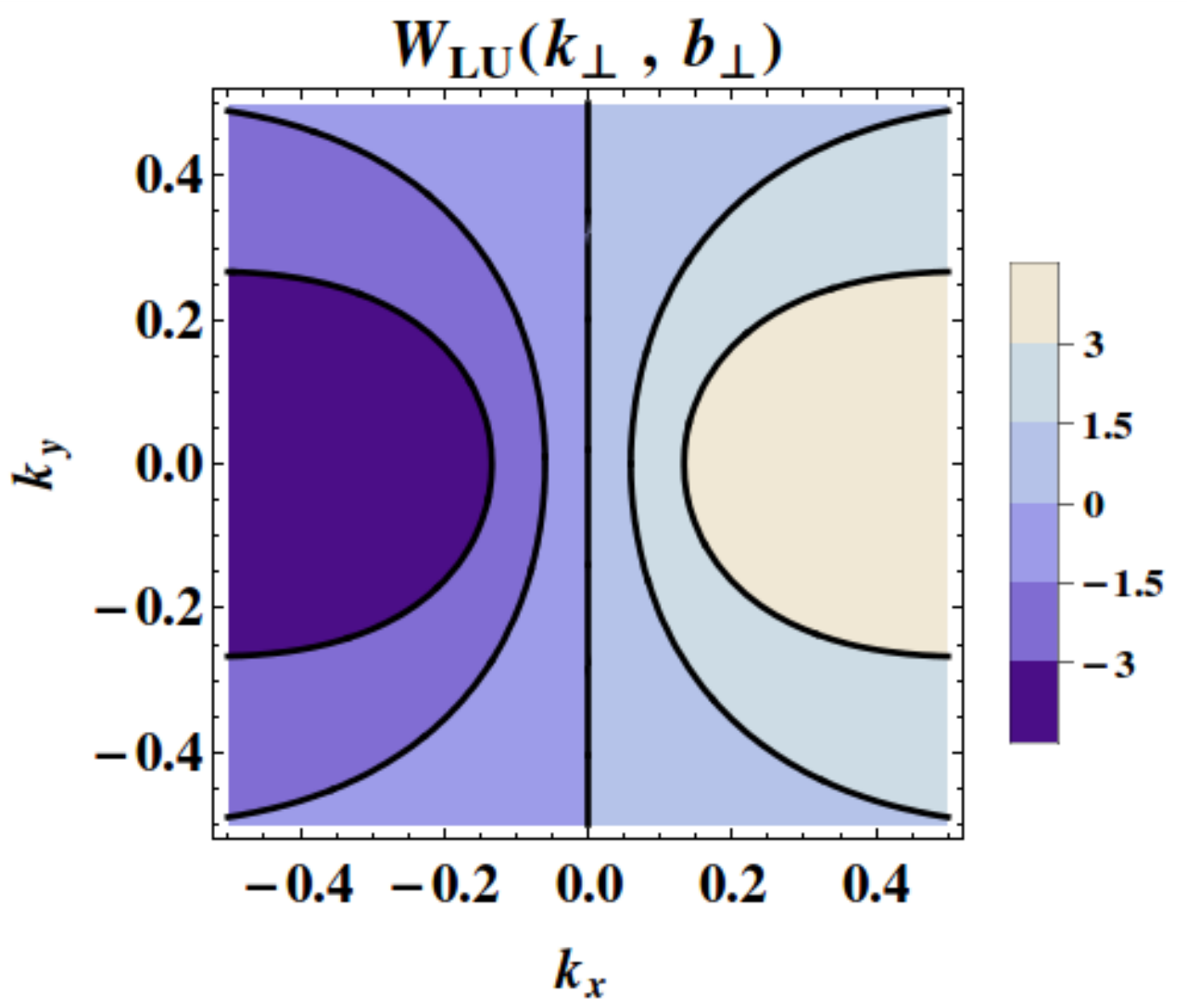}(e)\includegraphics[width=7.5cm,height=5.5cm]{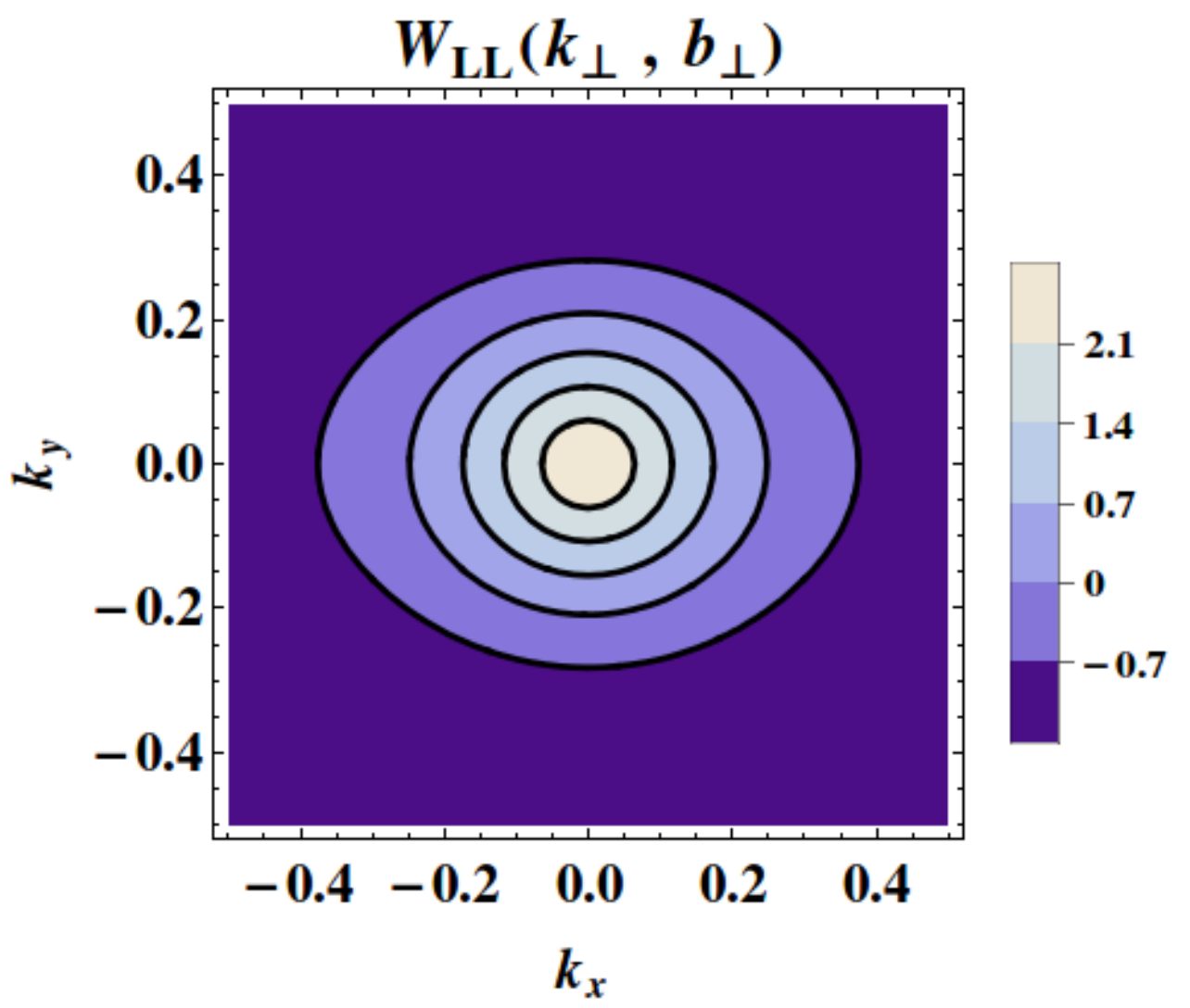}\\[2ex]
(c)\includegraphics[width=7.5cm,height=5.5cm]{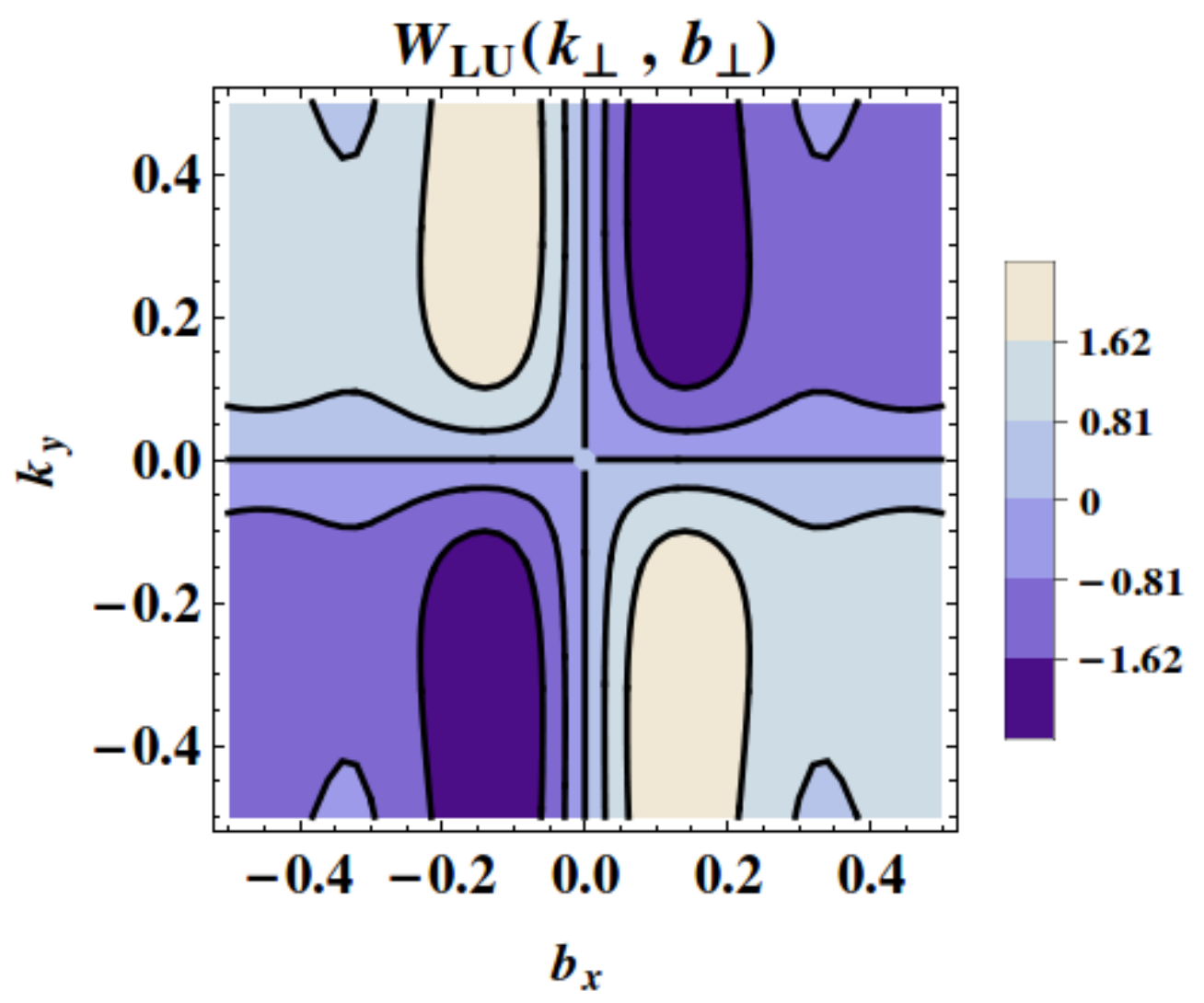}(f)\includegraphics[width=7.5cm,height=5.5cm]{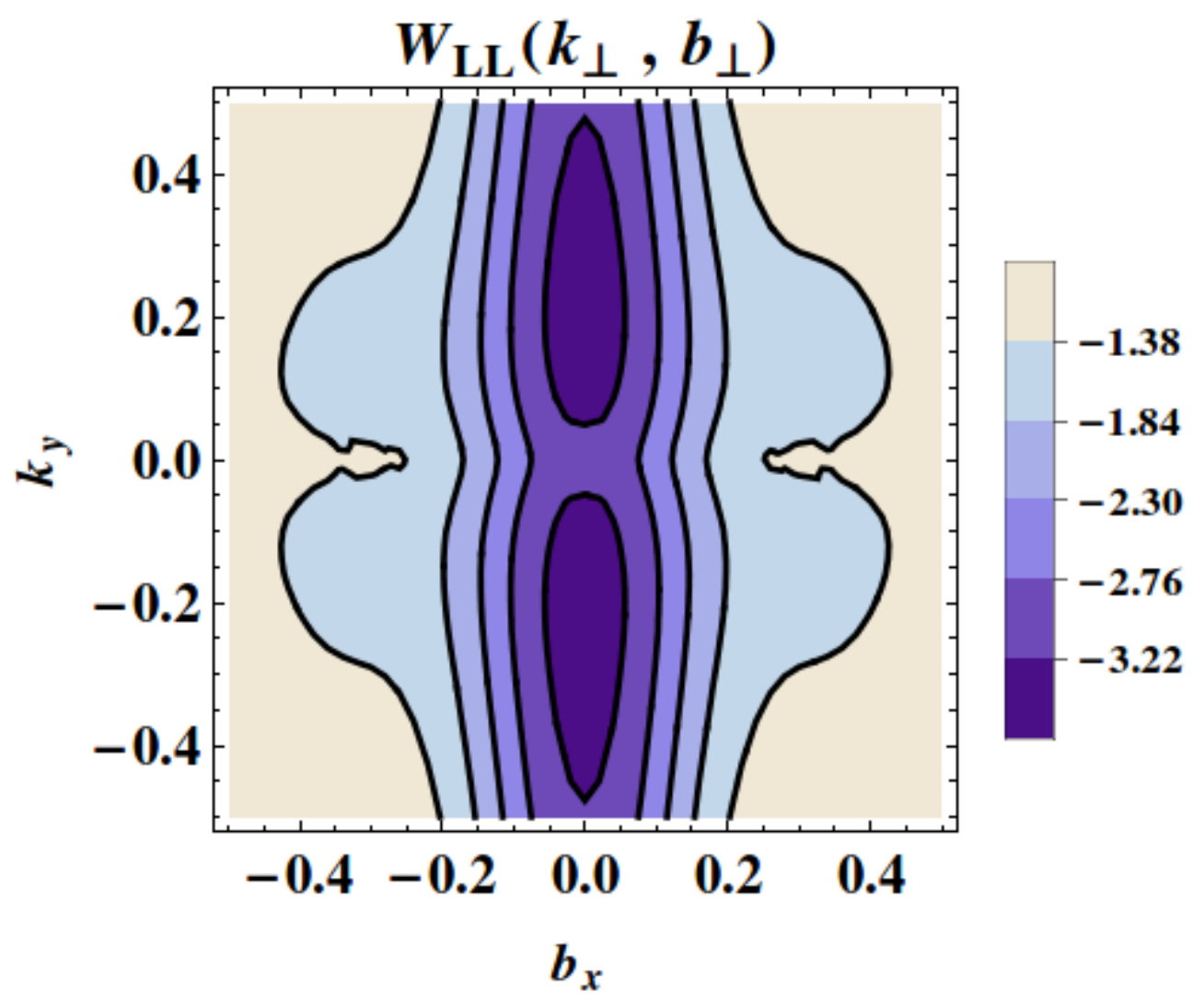}  
  \caption{Plot of  Wigner distributions $W_{LU}({\bs k}_\perp, {\bs b}_\perp)$ and 
  $W_{LL}({\bs k}_\perp, {\bs b}_\perp)$ at $\Delta_{max} = 20~\mathrm{GeV}$. The first row displays the two distributions
  in ${\bs b}_\perp$-space with ${\bs k}_\perp= 0.4~\mathrm{GeV} \,\hat{{\bs e}}_y$. The second row shows the two distributions
  in ${\bs k}_\perp$-space with ${\bs b}_\perp= 0.4~\mathrm{GeV}^{-1}\, \hat{{\bs e}}_y$.
  The last row represents the two distributions in mixed space.}  \label{wlu}
\end{figure}

\begin{figure}[h!]
(a)\includegraphics[width=7.5cm,height=5.5cm]{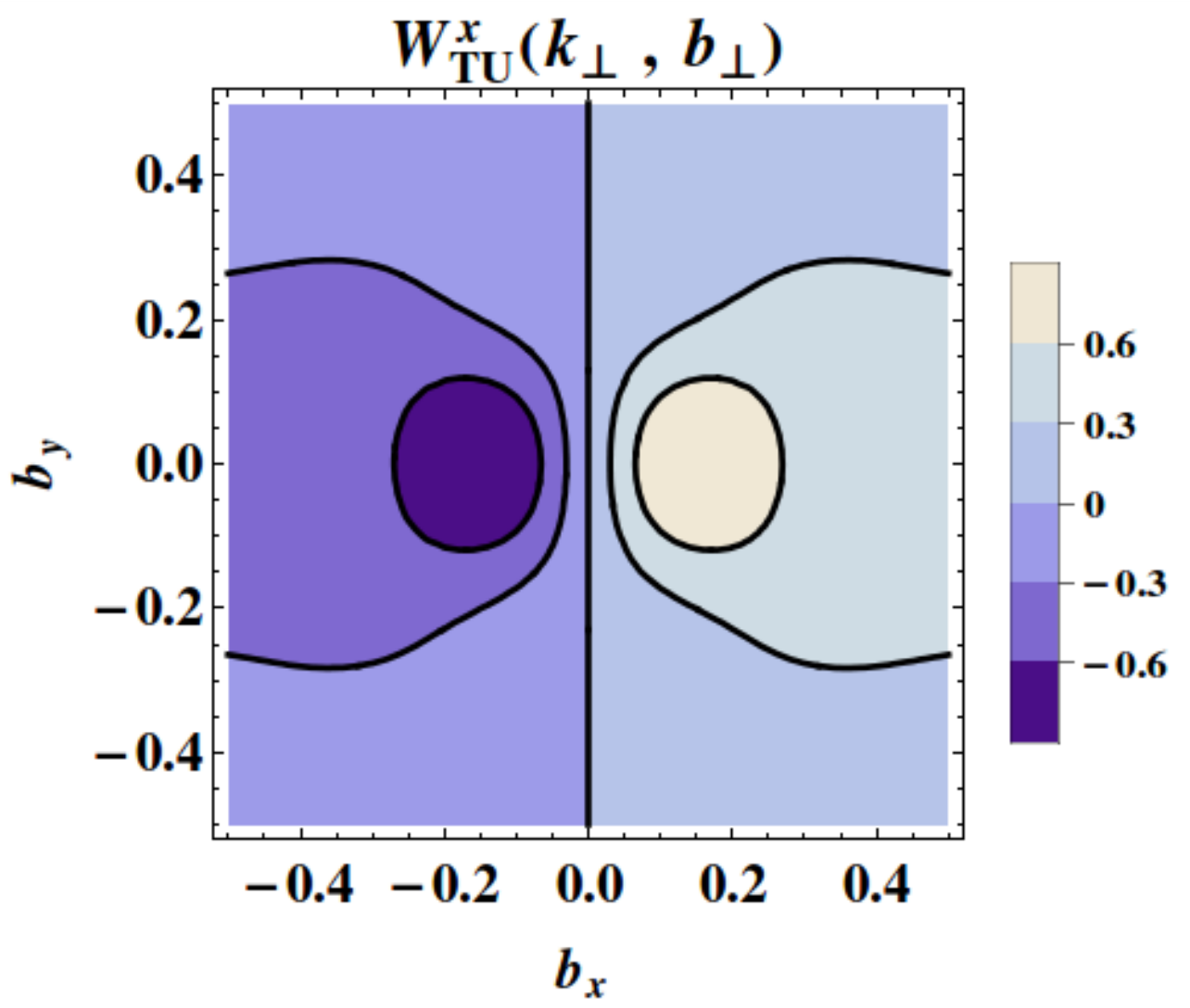}(d)\includegraphics[width=7.5cm,height=5.5cm]{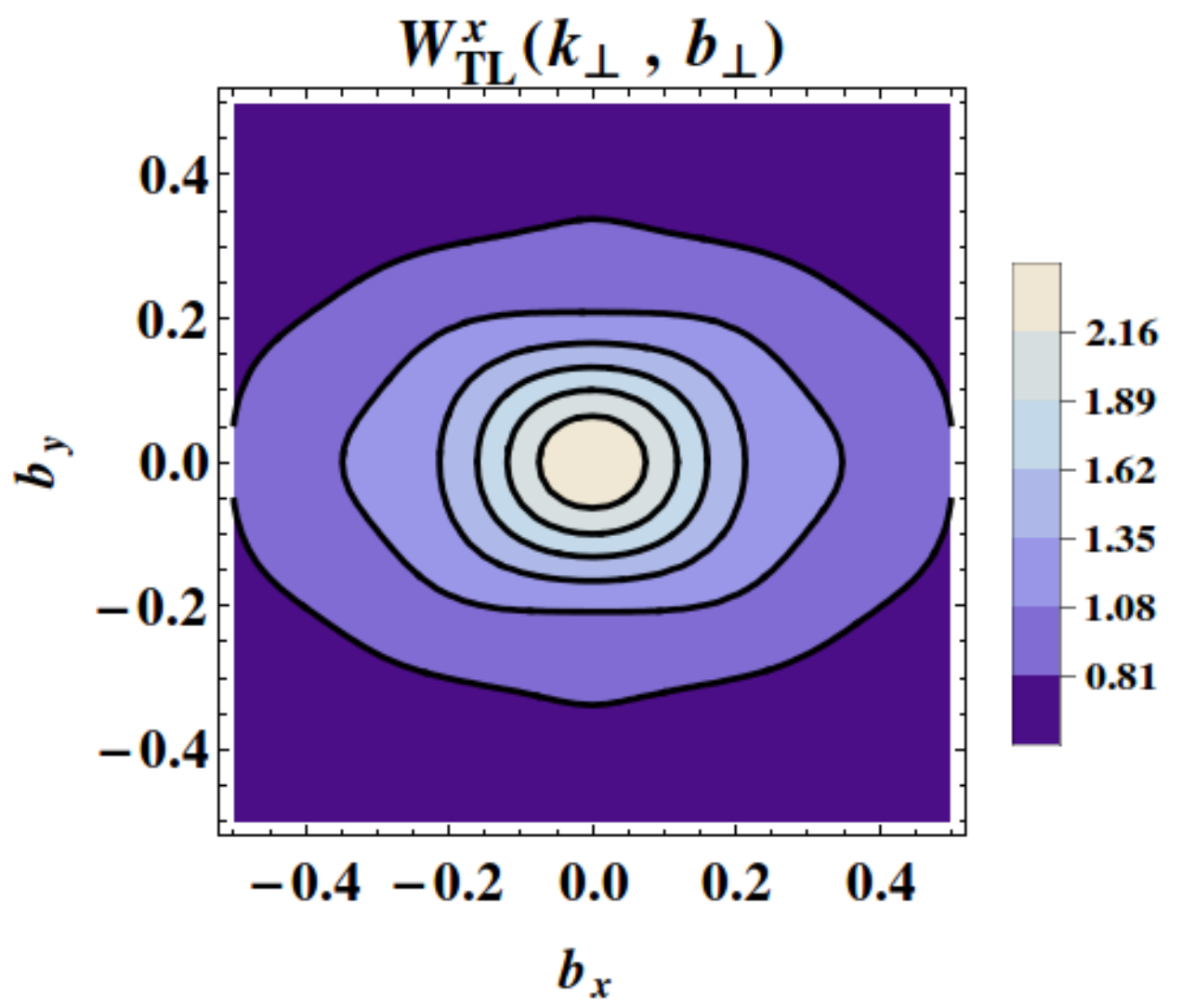}\\
(b)\includegraphics[width=7.5cm,height=5.5cm]{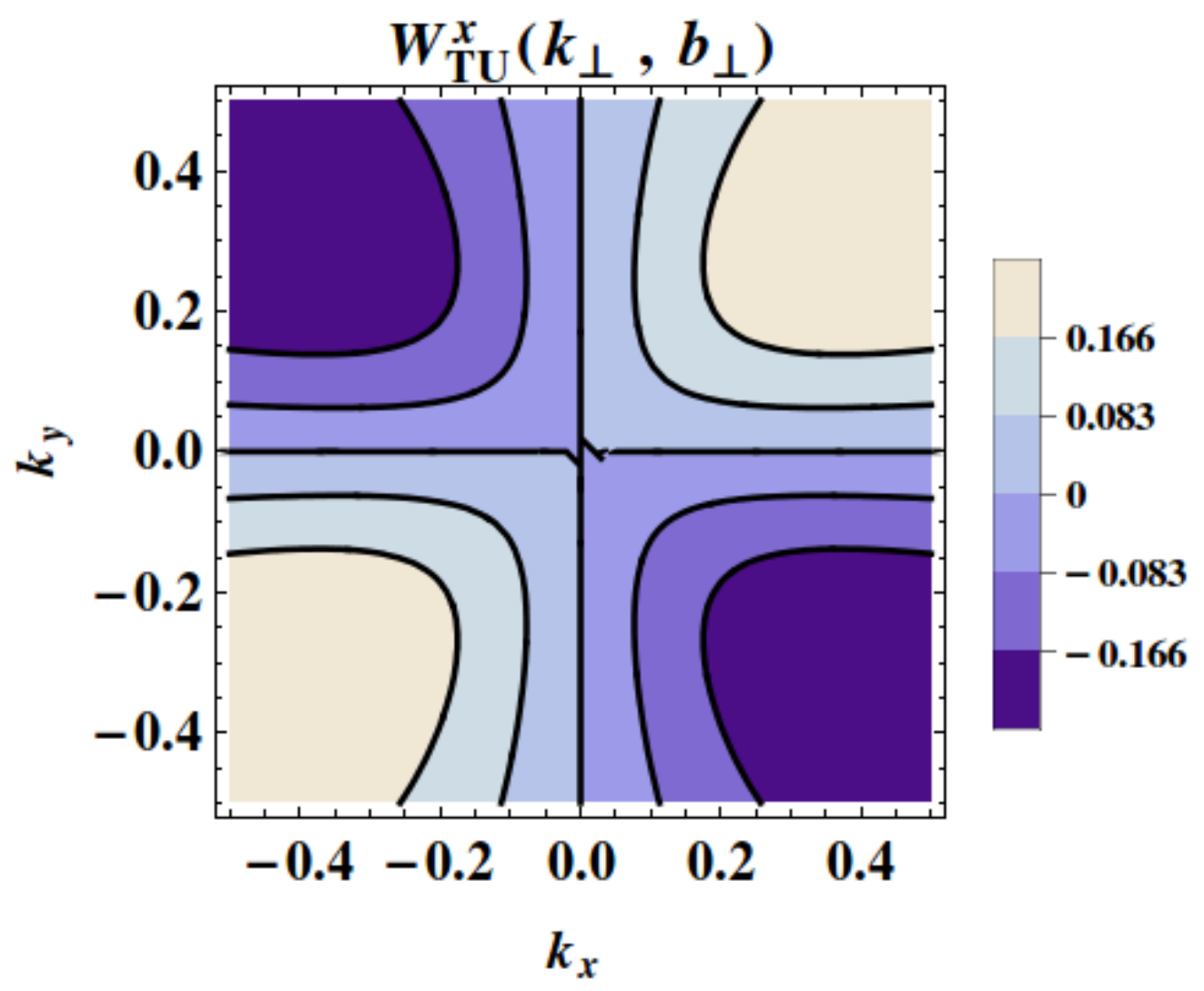}(e)\includegraphics[width=7.5cm,height=5.5cm]{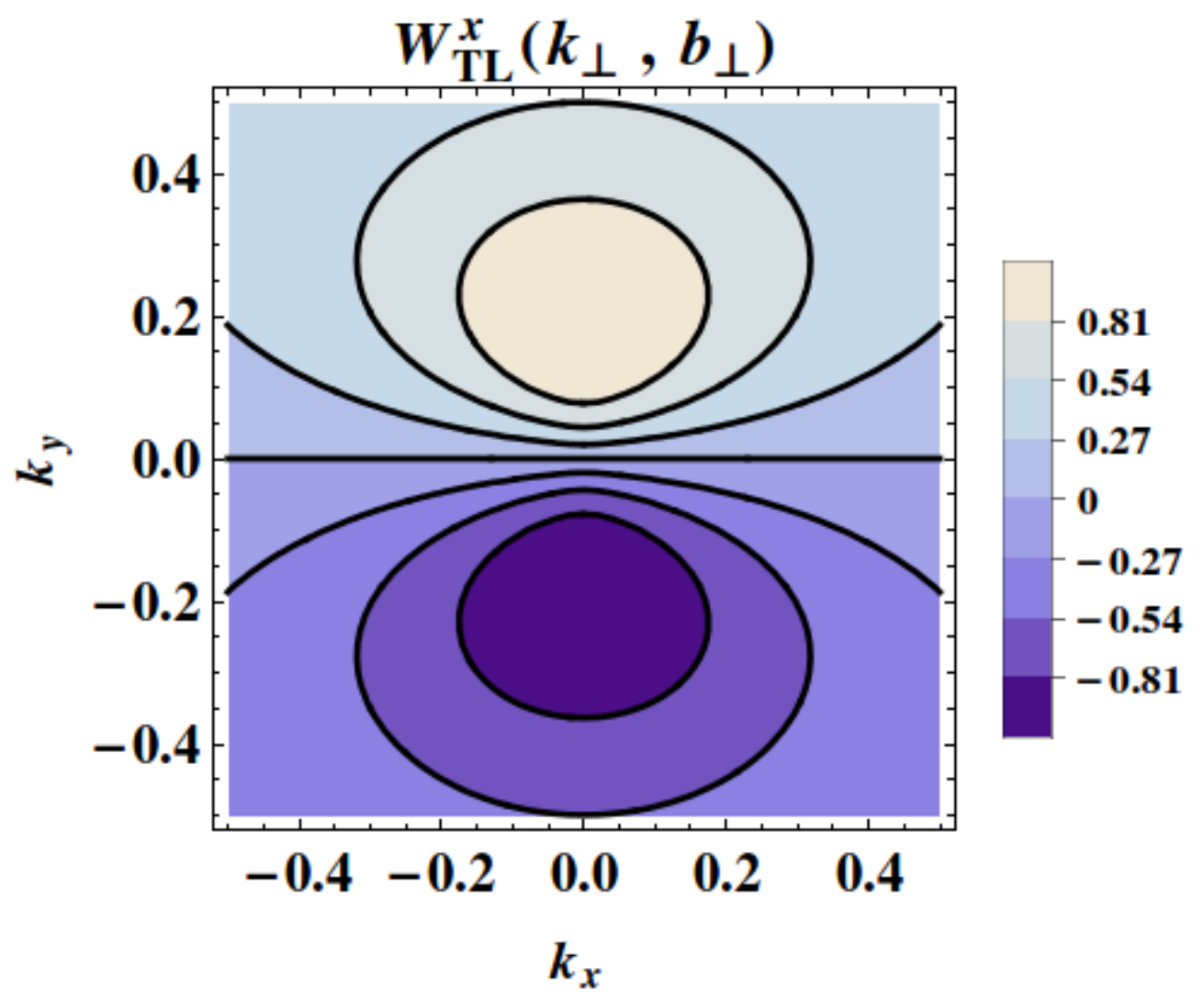}\\
(c)\includegraphics[width=7.5cm,height=5.5cm]{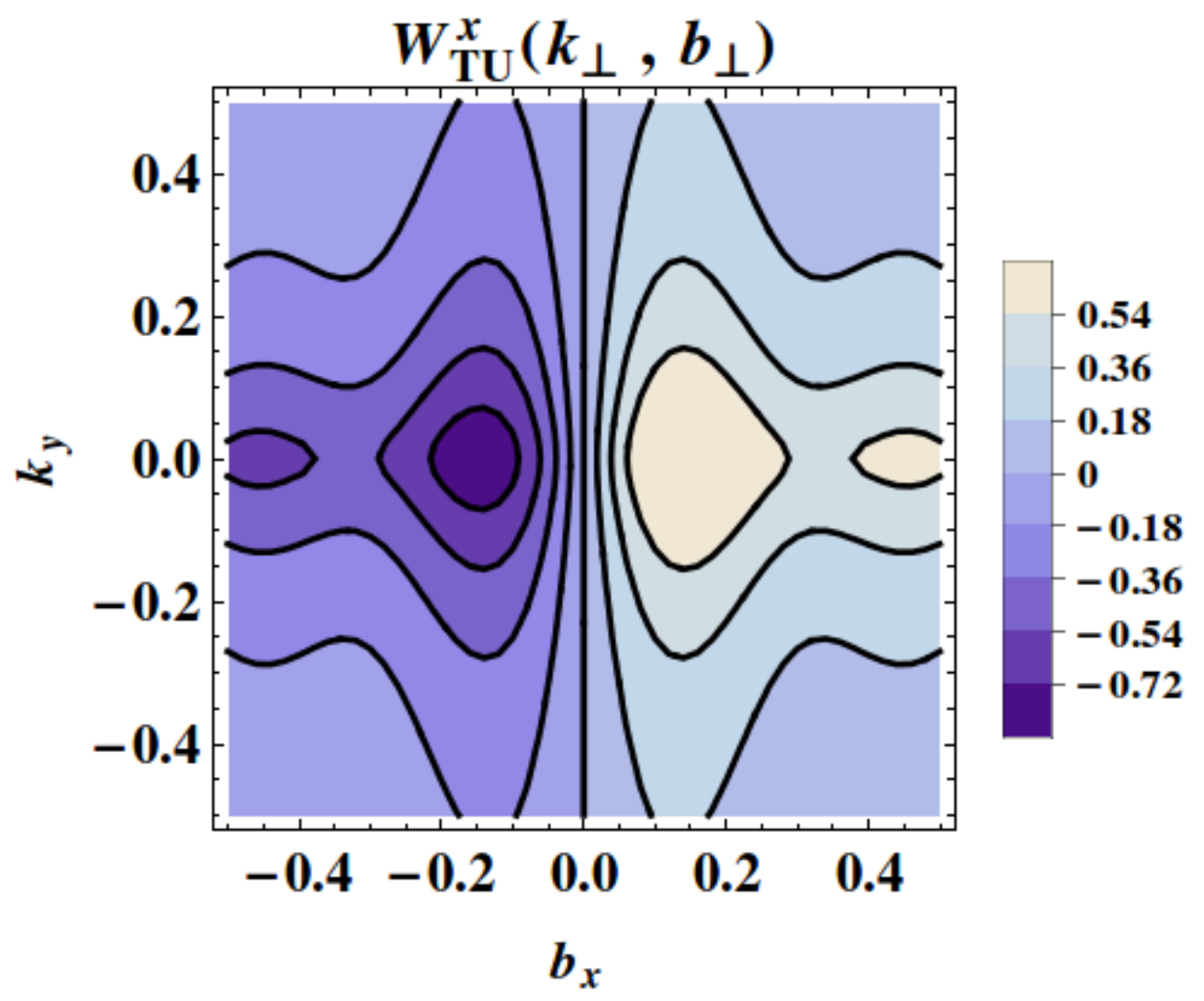}(f)\includegraphics[width=7.5cm,height=5.5cm]{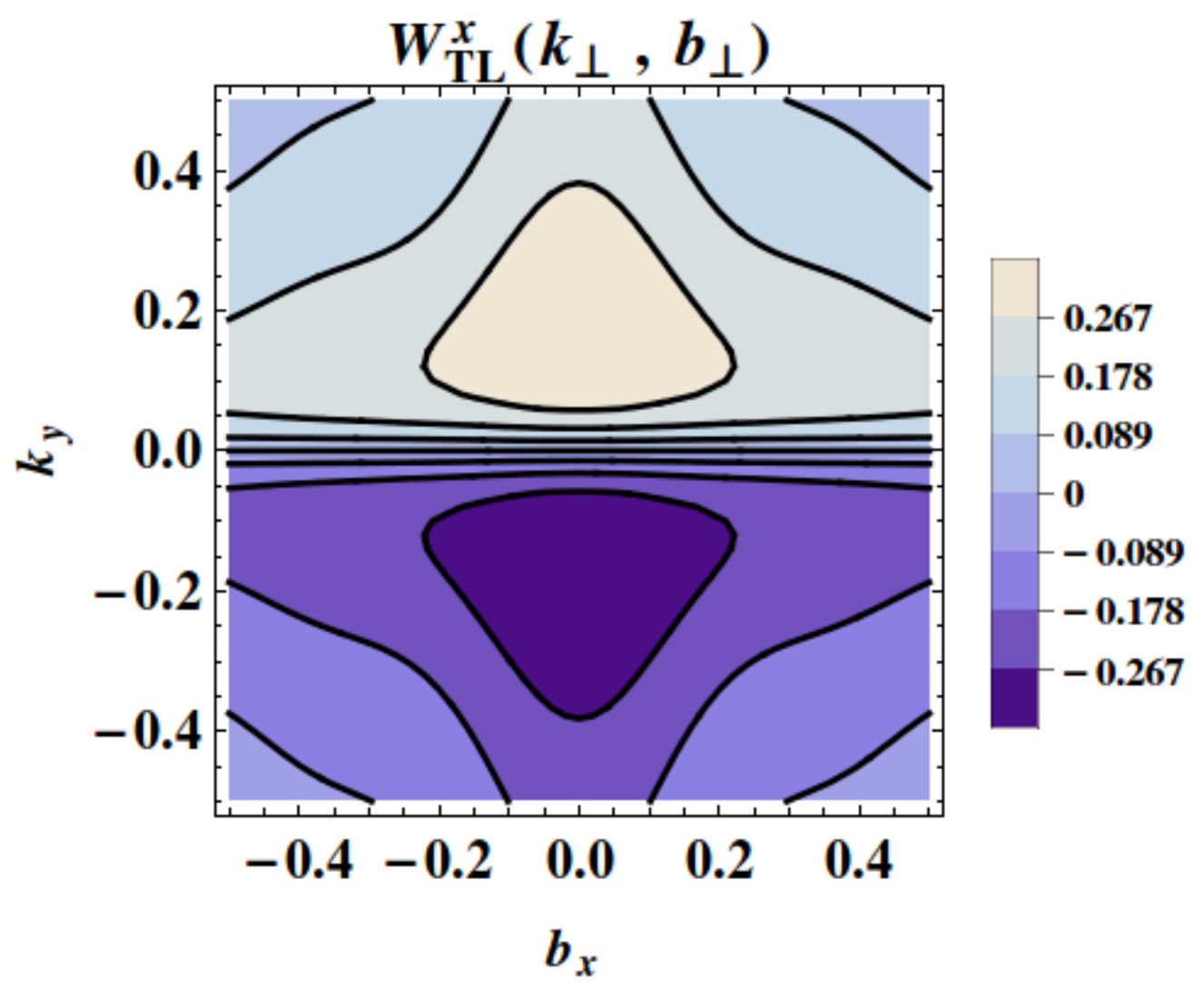}  
  \caption{Plot of  Wigner distributions $W_{T U}({\bs k}_\perp, {\bs b}_\perp)$ and 
  $W_{T L}({\bs k}_\perp, {\bs b}_\perp)$ at $\Delta_{max} = 20~\mathrm{GeV}$. The first row displays the two distributions
  in ${\bs b}_\perp$-space with ${\bs k}_\perp= 0.4~\mathrm{GeV} \,\hat{{\bs e}}_y$. The second row shows the two distributions  in ${\bs k}_\perp$-space with ${\bs b}_\perp= 0.4~\mathrm{GeV}^{-1}\, \hat{{\bs e}}_y$.
  The last row represents the two distributions in mixed space.} 
 \label{wtu}
\end{figure}

\begin{figure}[h]
(a)\includegraphics[width=7.5cm,height=5.5cm]{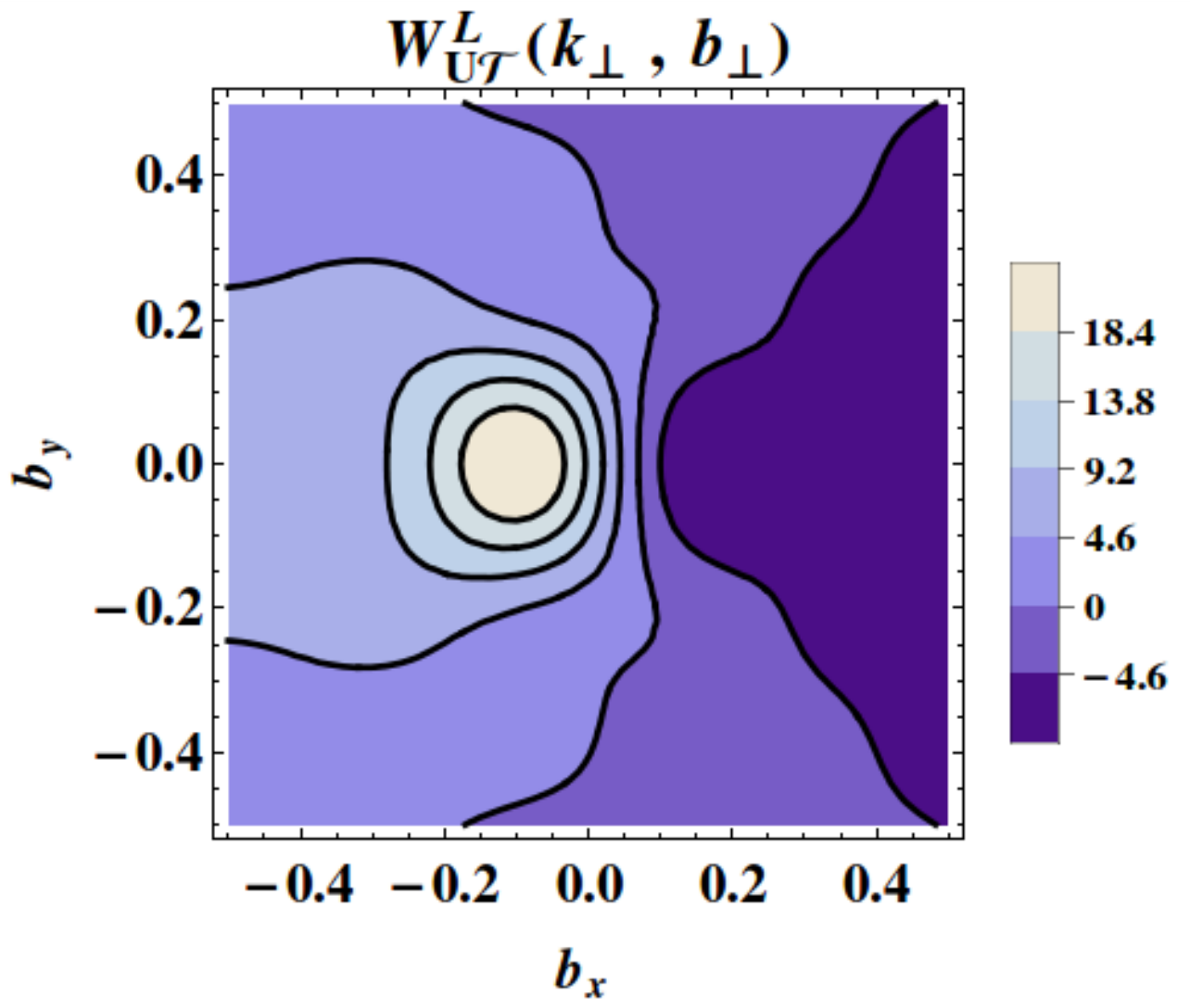}(d)\includegraphics[width=7.5cm,height=5.5cm]{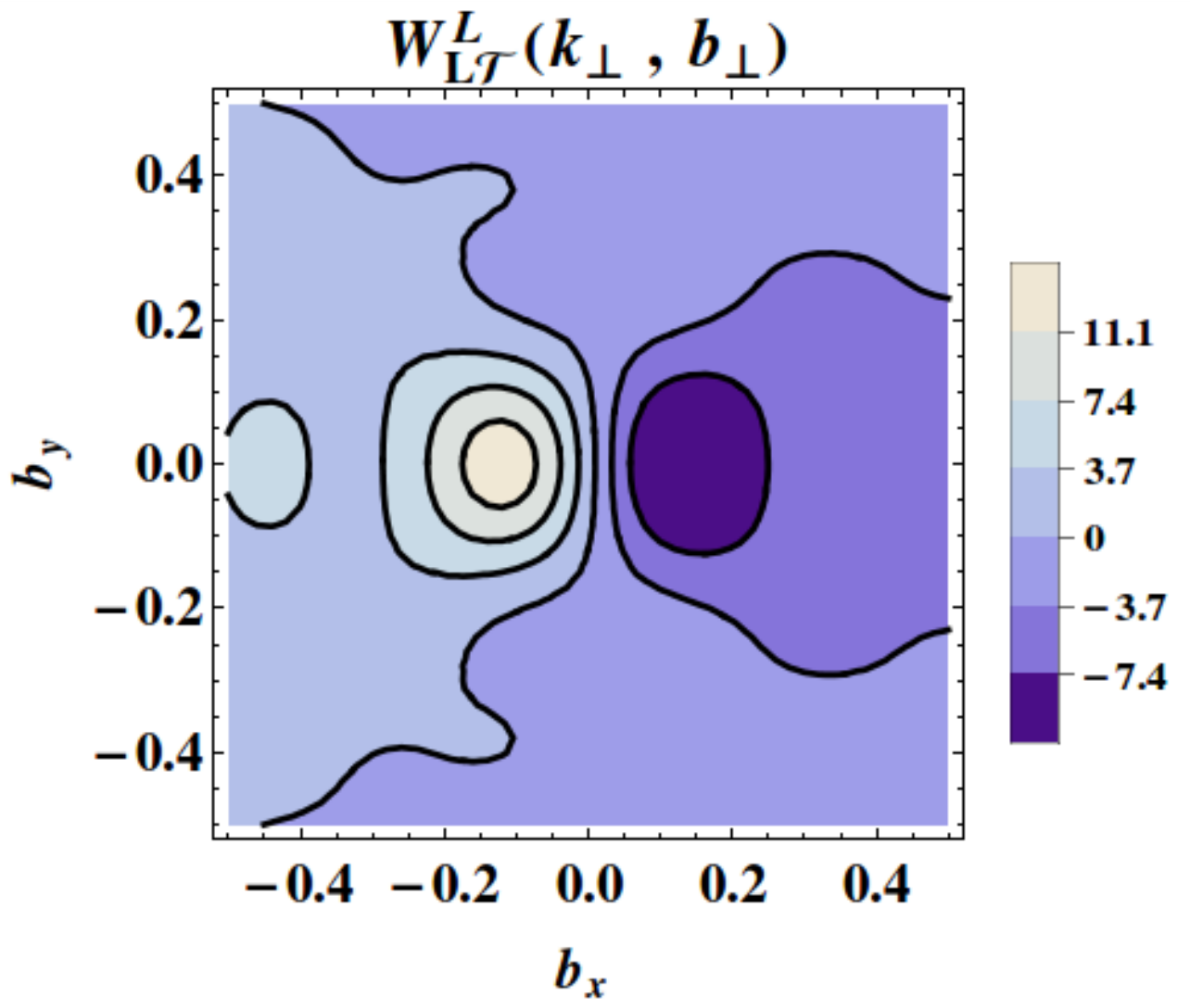}\\
(b)\includegraphics[width=7.5cm,height=5.5cm]{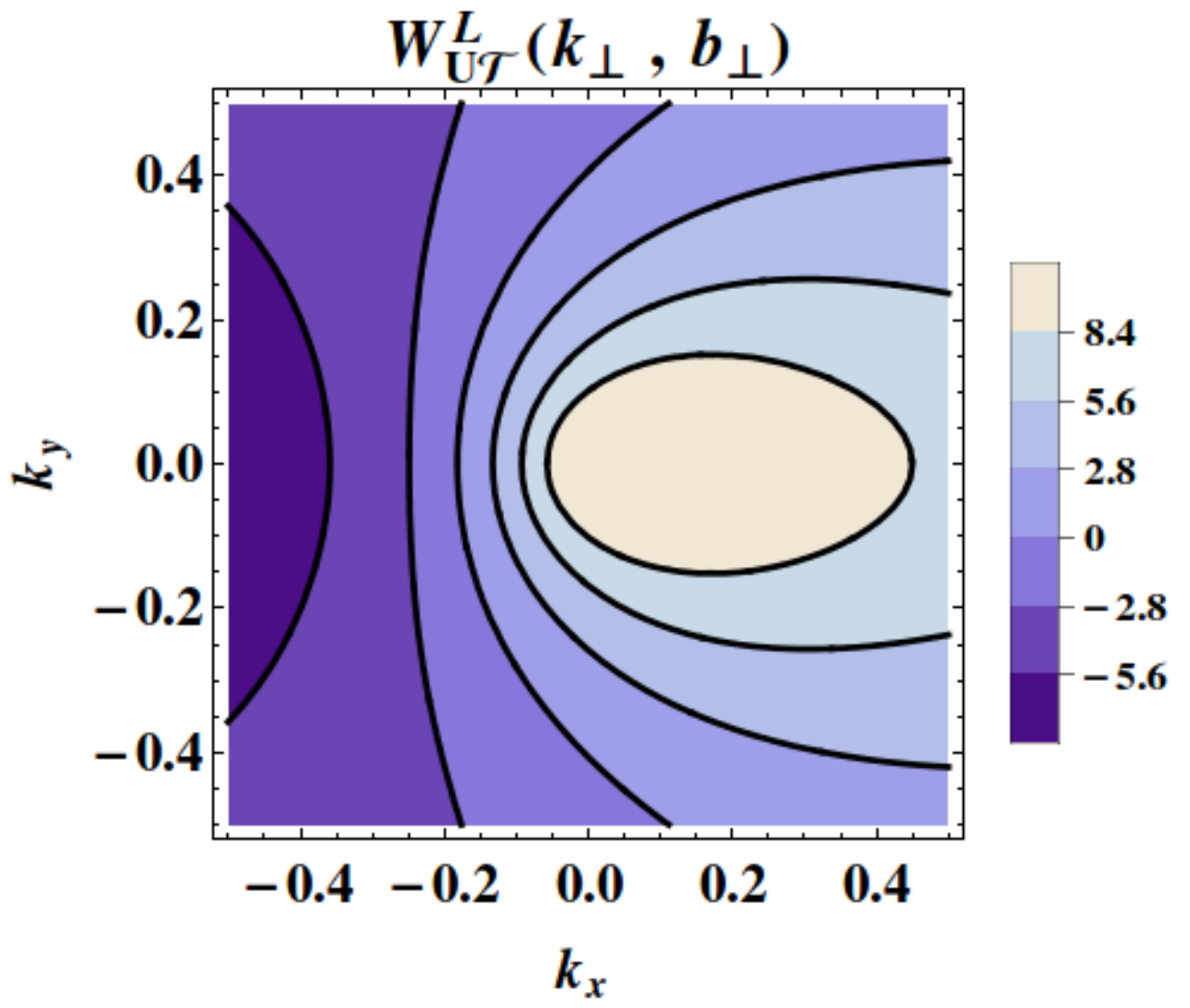}(e)\includegraphics[width=7.5cm,height=5.5cm]{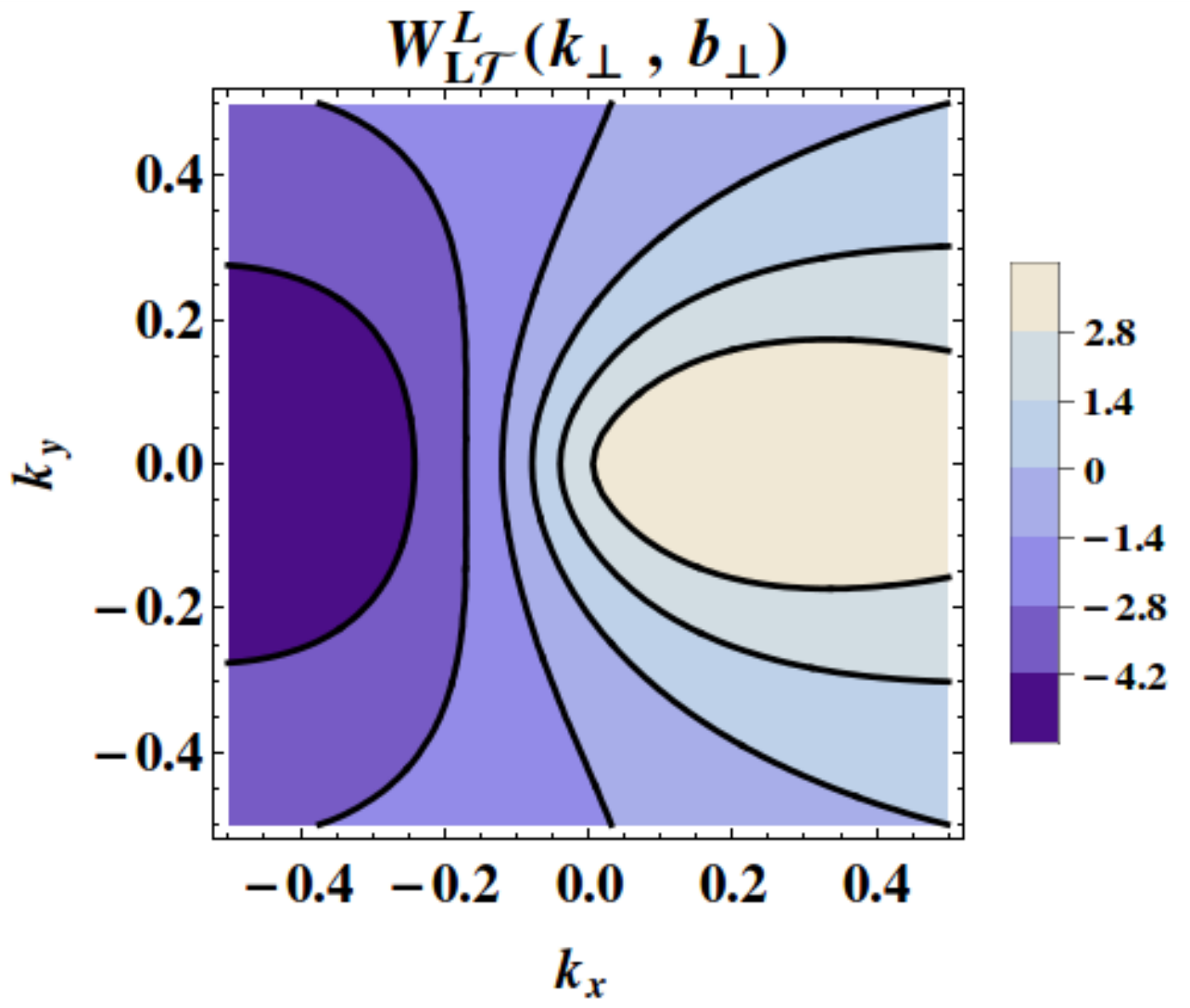}\\
(c)\includegraphics[width=7.5cm,height=5.5cm]{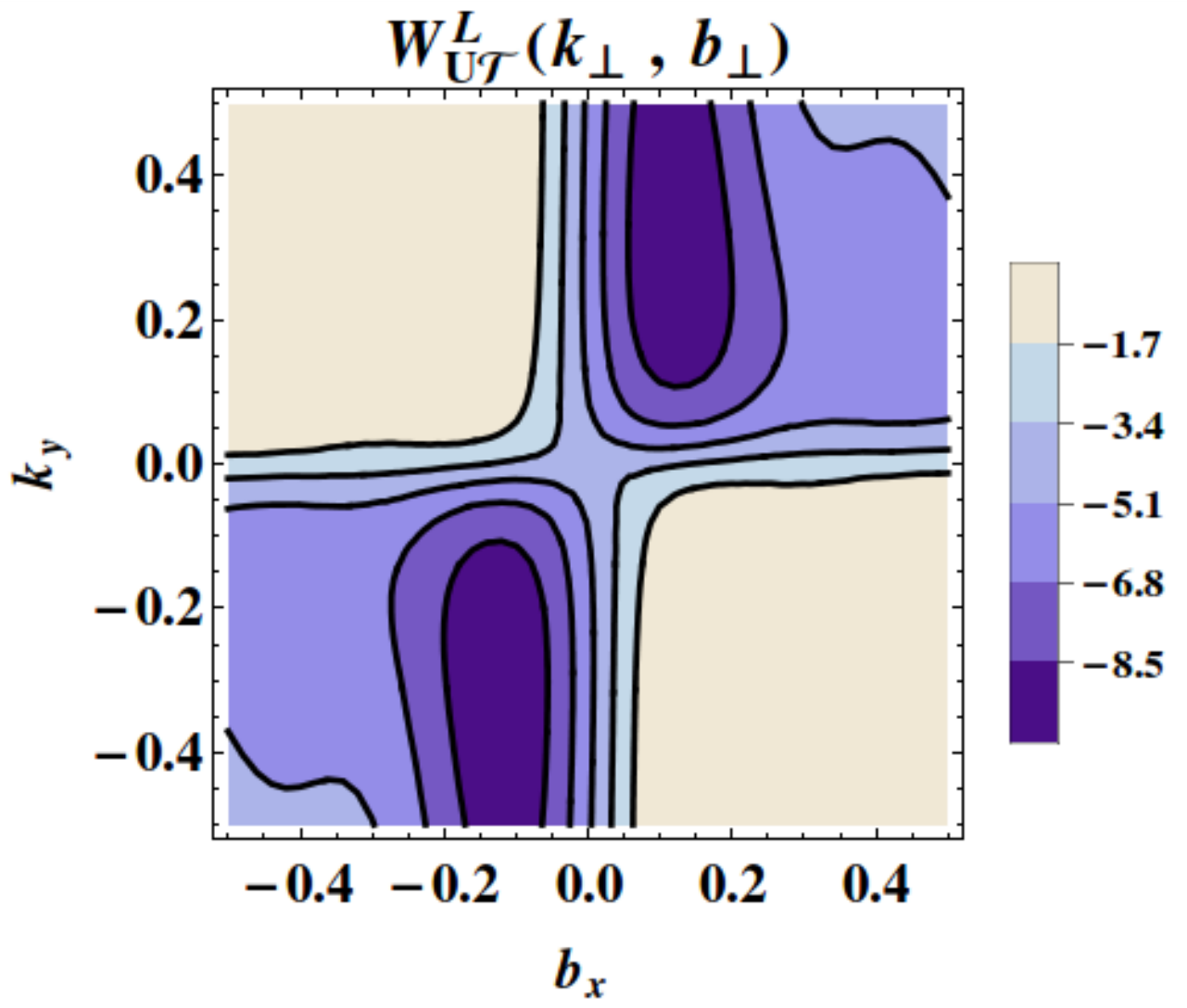}(f)\includegraphics[width=7.5cm,height=5.5cm]{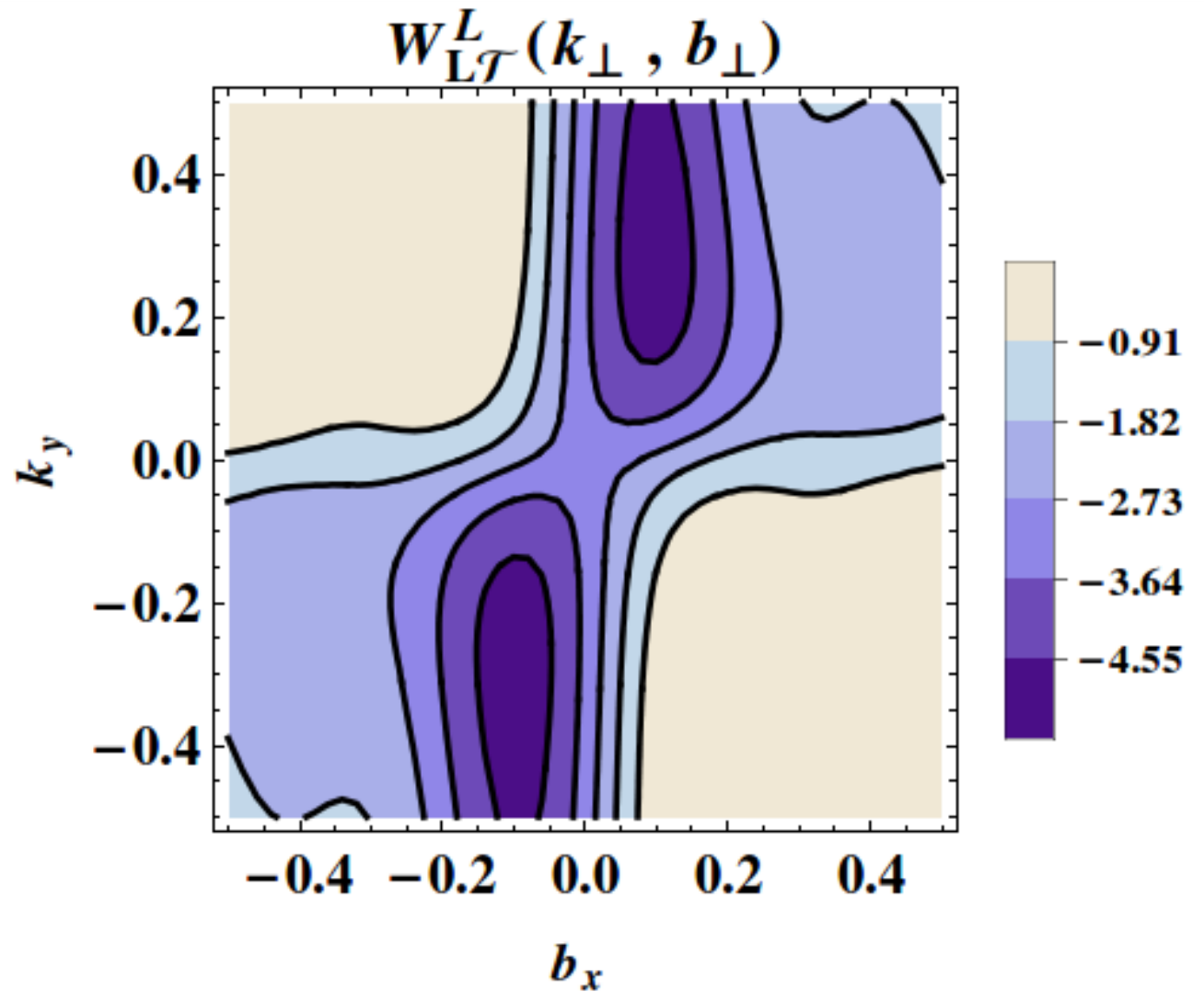}  
  \caption{Plot of  Wigner distributions $W^{L}_{U\T}({\bs k}_\perp, {\bs b}_\perp)$ and 
  $W^{L}_{L\T}({\bs k}_\perp, {\bs b}_\perp)$ at $\Delta_{max} = 20~\mathrm{GeV}$. The first row displays the two distributions
  in ${\bs b}_\perp$-space with ${\bs k}_\perp= 0.4~\mathrm{GeV} \,\hat{{\bs e}}_y$. The second row shows the two distributions
  in ${\bs k}_\perp$-space with ${\bs b}_\perp= 0.4~\mathrm{GeV}^{-1}\, \hat{{\bs e}}_y$. The last row represents the two distributions in mixed space.}
 \label{wut}
\end{figure}

\begin{figure}[h]
(a)\includegraphics[width=7.5cm,height=5.5cm]{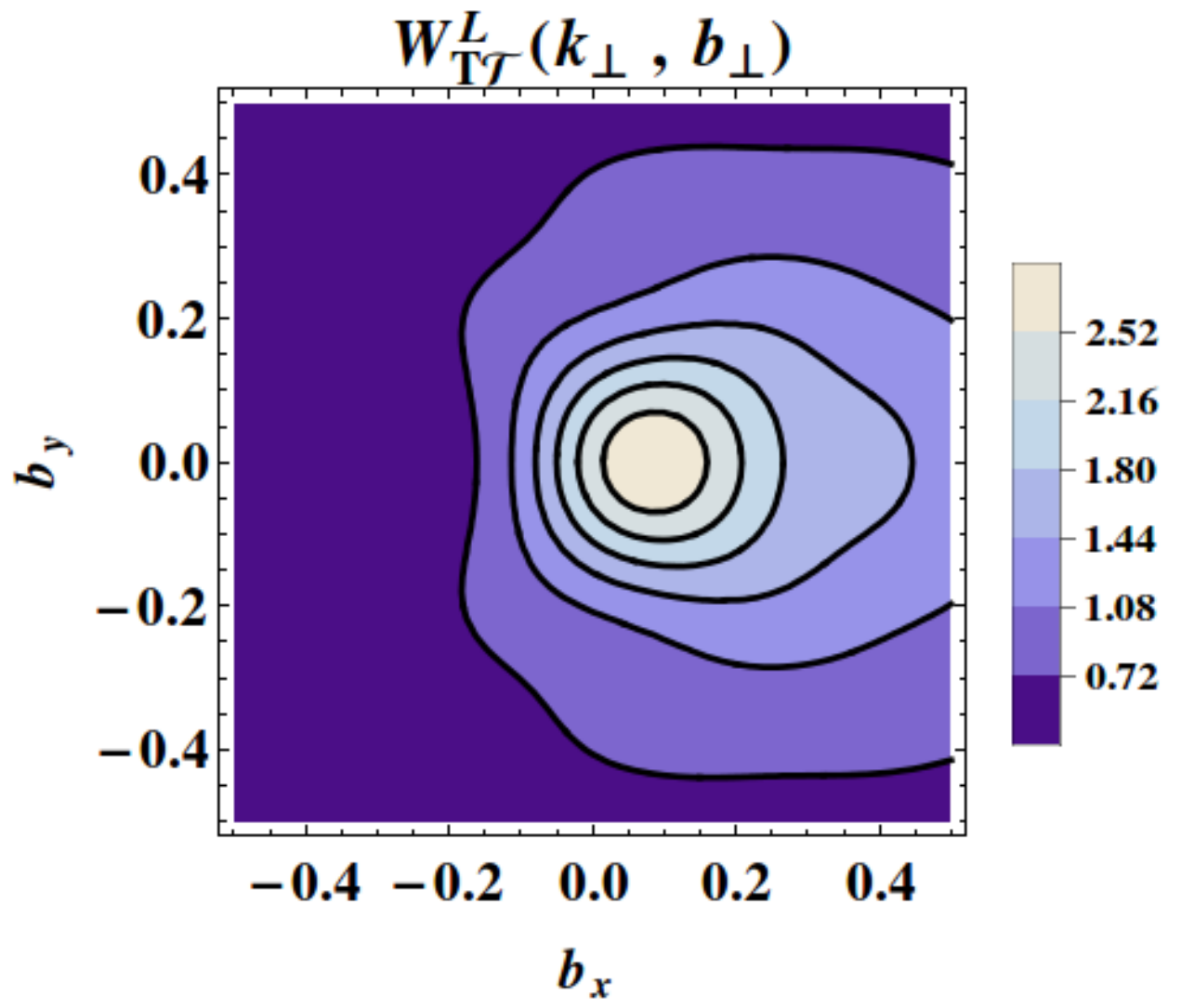}(d)\includegraphics[width=7.5cm,height=5.5cm]{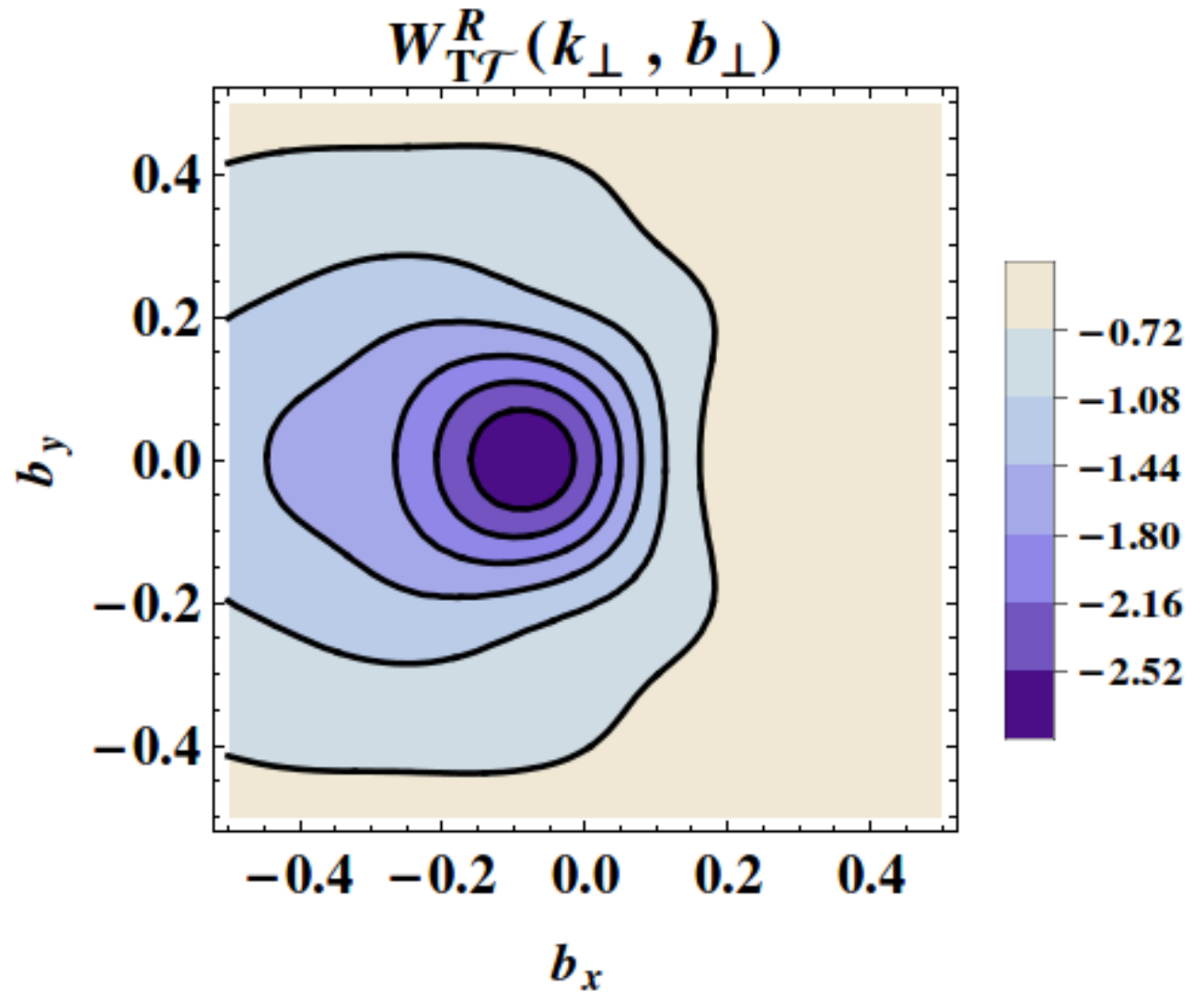}\\
(b)\includegraphics[width=7.5cm,height=5.5cm]{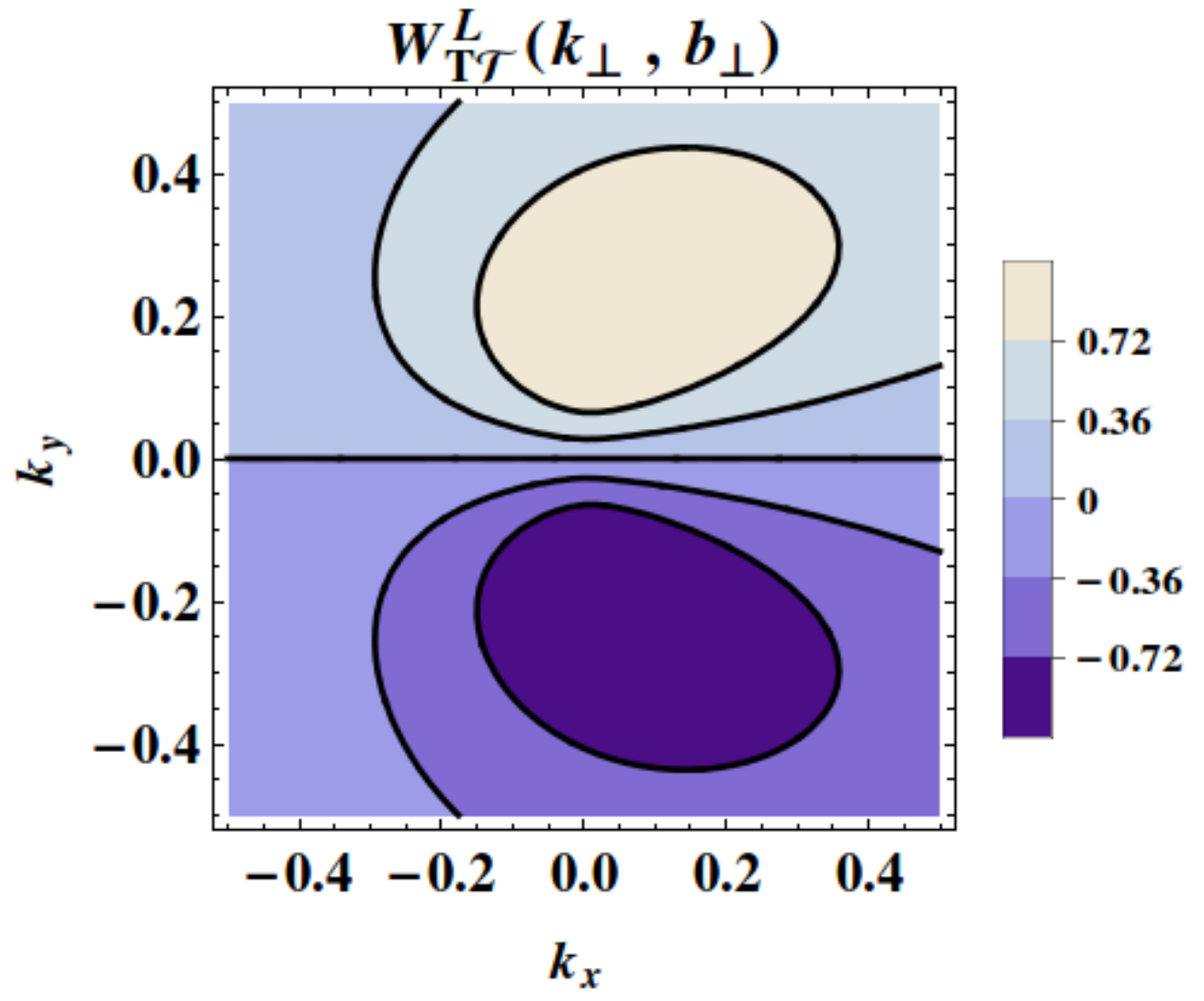}(e)\includegraphics[width=7.5cm,height=5.5cm]{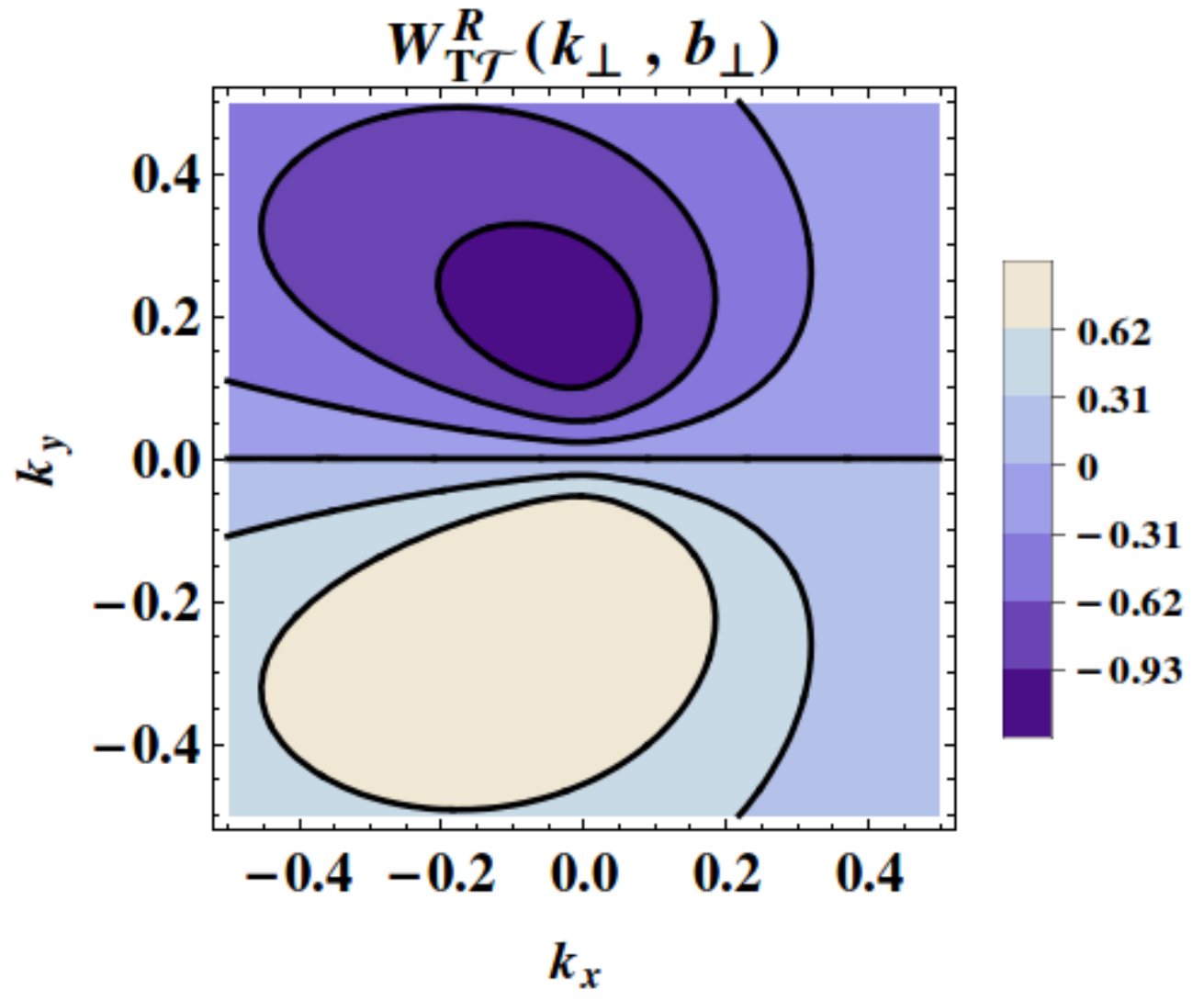}\\(c)\includegraphics[width=7.5cm,height=5.5cm]{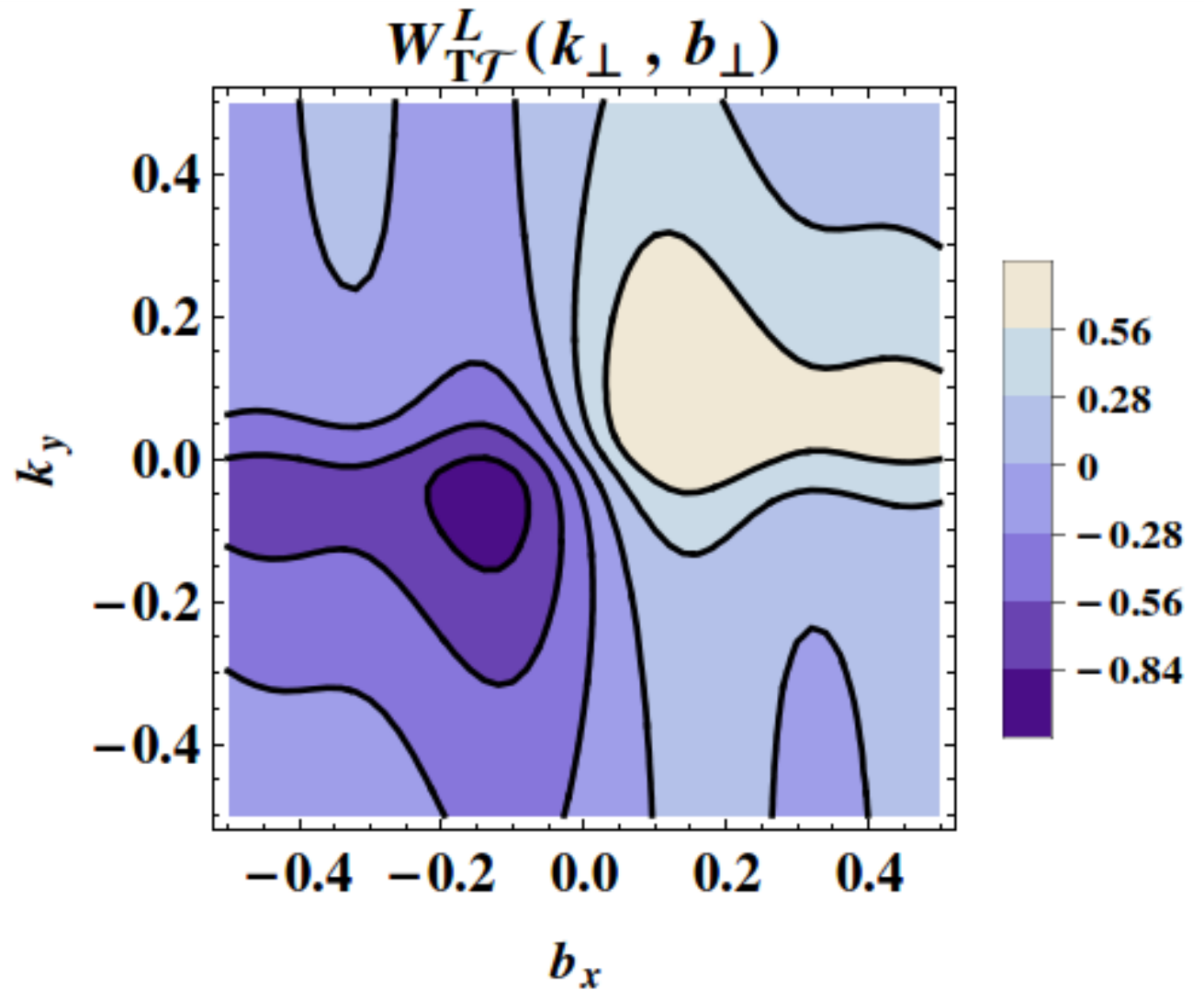}(f)\includegraphics[width=7.5cm,height=5.5cm]{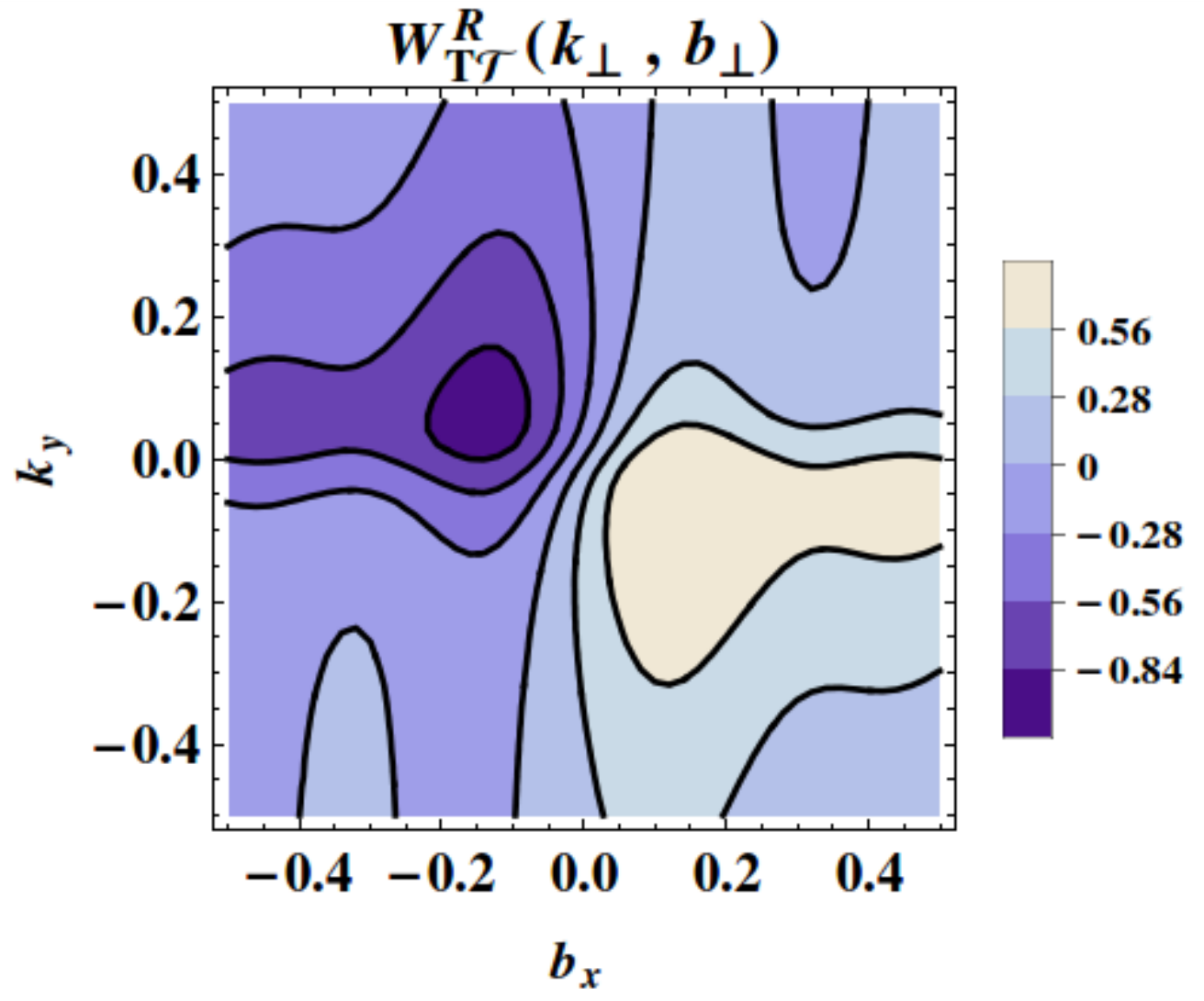}  
  \caption{Plot of Wigner distributions $W^{L}_{T \T}({\bs k}_\perp, {\bs b}_\perp)$ and 
  $W^{R}_{T \T}({\bs k}_\perp, {\bs b}_\perp)$ at $\Delta_{max} = 20~\mathrm{GeV}$. The first row displays the two distributions
  in ${\bs b}_\perp$-space with ${\bs k}_\perp= 0.4~\mathrm{GeV} \,\hat{{\bs e}}_y$. The second row shows the two distributions
  in ${\bs k}_\perp$ space with ${\bs b}_\perp= 0.4~\mathrm{GeV}^{-1}\, \hat{{\bs e}}_y$. The last row represents the two distributions in mixed space.}
 \label{wtt}
\end{figure}

The Wigner distributions for different gluon and target polarizations are shown in contour plots of 
Figs. \ref{wuu} - \ref{wtt}. We have plotted these in the ${\bs b}_\perp$ plane by fixing  $k _\perp$ = 0.4 GeV $\hat{e_j}$.
For the plot in ${\bs k_\perp}$ plane we have fixed $b_\perp$ = 0.4 GeV $\hat{e_j}$. Finally we have integrated over $b_y$ and $k_x$ and showed plots in the mixed space of $k_y$ and $b_x$.  $W_{UU}$ is shown in Fig. \ref{wuu} (a)-(c), this describes the distribution of  unpolarized gluons in an unpolarized target.  $W_{UU}$ has a positive peak at the center both in ${\bs b}_\perp$ space and in ${\bs k}_\perp$ space. The distribution is spread more in the $x$ direction compared to $y$ direction,  this means that it is more probable to find a gluon with large $b_x/k_x$. This asymmetry is introduced by the choice of ${\bs k}_\perp/{\bs b}_\perp$ in the $y$ direction and is similar to that observed for quarks in Light-cone CQM \cite{Lorce11}. The gluon here is more likely to have ${\bs k}_\perp \perp {\bs b}_\perp$ rather than ${\bs k_\perp}$ parallel to ${\bs b_\perp}$. This asymmetry represents a correlation between ${\bs b}_\perp$ and ${\bs k}_\perp$, however this is not related to any confining interaction, as there is no confinement potential in our perturbative model. The top-bottom symmetry of the distributions in ${\bs b}_\perp$ and ${\bs k}_\perp$ space is related to the  absence of the gauge link term in our study, as discussed in \cite{Lorce11}. The mixed space plots represent   correlations between $b_x$ and $k_y$.  It is to be noted that the behaviour of 
$W_{UU}$ near $b_\perp=0$ is governed by the relative dominance  of the $k_\perp^2$ and $\Delta_\perp^2 (1-x)^2$ terms in the numerator. As $\Delta_{max}$ increases, the second term dominates over the first, as a result, the peak at $b_\perp=0$ becomes positive. As we increase $\Delta_{max}$ beyond $20$ GeV, the behavior of all Wigner distributions is independent of the cutoff.     

In Figs. \ref{wuu} (d)-(f), we have plotted $W_{UL}$ and in Figs. \ref{wlu} (a)-(c) we have shown $W_{LU}$. These distributions do not have a TMD limit. They describe the distribution of longitudinally polarized  gluon in unpolarized target and unpolarized gluon in the longitudinally polarized target, respectively. $W_{LU}$ is related to the gluon OAM and $W_{UL}$ is related to the spin-orbit correlation of the gluon \cite{Lorce11,Hatta12}. Both gluon OAM and spin-orbit correlation in this model have been discussed in \cite{Mukherjee15, Kanazawa14}. In a later section, we have discussed the small  $x$ limit of the quark and gluon helicity and OAM. $W_{LU}$ represents a distortion in impact parameter space due to non-zero OAM and $W_{UL}$ represents a distortion due to spin-orbit coupling. These distortions in our model are dipolar in nature. In mixed space we observe a quadrupole behaviour.  The spin of the gluon is preferably anti-aligned with the OAM.  In ${\bs k}_\perp$ space, the correlations occur over a broad region than in ${\bs b}_\perp$-space. As most phenomenological models do not have gluons, these are the first model study of the gluon Wigner distributions  integrated over $x$. Figs. \ref{wlu} (d)-(f) show the plots of  $W_{LL}$, which is the distribution of longitudinally polarized gluon in a longitudinally polarized target. These are also peaked at the center of ${\bs b}_\perp$ and ${\bs k}_\perp$  space and spread along the x direction;  indicating that ${\bs b}_\perp \perp {\bs k}_\perp$ is favored for longitudinally polarized gluons. In the mixed space, two peaks are observed. $W_{LL}$ is negative in the outer region of the ${\bs b}_\perp$ and ${\bs k}_\perp$ space.

Figs.\ref{wtu} (a)-(c) show plots of $W^x_{TU}$, which is the distribution of unpolarized gluons in the  transversely polarized target. Polarization of the target is taken to be in the $x$ direction.  In ${\bs b}_\perp$ space this shows a dipole nature, and the distribution is spread towards larger  $b_x$. In ${\bs k}_\perp$ space a quadrupole nature is observed. The correlation is positive when both $k_x$ and $k_y$ have the same sign. One positive and one negative peak are seen in the mixed space. Figs. \ref{wtu} (d)-(f) show plots of $W^x_{TL}$, which is the distribution of longitudinally polarized gluons in a transversely polarized target.The polarization of the target is taken to be in the x direction. The asymmetry is introduced here by the polarization of the transversely polarized target. The distortion of longitudinally polarized  gluon in a transversely polarized target is sharply peaked at the centre of the ${\bs b}_\perp$ space, spread more in the direction of $b_x$, showing that ${\bs k}_\perp \perp {\bs b}_\perp$ configuration is more favored. In ${\bs k}_\perp$ space the distribution vanishes when ${\bs k}_\perp \perp {\bs b}_\perp$, is positive in the upper half plane and negative in the lower half.   Similar behavior is observed in mixed space.  Figs. \ref{wut} (a)-(c) show plots of $W^L_{UT}$ and Figs. \ref{wut} (d)-(f) show plots of $ W^L_{LT}$, the definitions of these distributions are given earlier. These probe linearly polarized  gluons in unpolarized and longitudinally polarized target, respectively. Qualitatively, the behavior  of both are similar, in ${\bs b}_\perp$ space, a dipole nature is observed, with a negative peak along positive $b_x$ axis. In ${\bs k}_\perp$ space also a dipolar nature is seen, but with a broad positive  peak in the positive $k_x$  region. There is a left-right asymmetry in the  ${\bs k}_\perp$ space, which is related to the linear polarization of the gluon, as it is present when the target is unpolarized as well as when it is longitudinally polarized.  In mixed space two negative peaks are seen. Figs. \ref{wtt} (a)-(c) show plots of $W^L_{TT}$ and Figs. \ref{wtt} (d)-(f) show $W^R_{TT}$. Both are similar but have opposite polarity. In ${\bs b}_\perp$ space the distribution is asymmetric, $W^L_{TT}$ is shifted towards  positive $b_x$ whereas $W^R_{TT}$ is shifted towards negative $b_x$. The peak is also inverted. In ${\bs k}_\perp $ space, a dipolar nature is seen and the distribution is broad. The distribution is spread over a wide region in the mixed space. It will be interesting to see if these 
behaviors are also observed in the gluon Wigner distributions in a phenomenological model, also  
whether it is possible to experimentally access them. Both these aspects are beyond the scope of the present work.  


  
The GTMDs are related to various TMDs in the forward limit ${\bs \Delta_\perp}=0$. The GTMDs are related to the GPDs in impact parameter space upon
integration over ${\bs k_\perp}$. The relation of the different GTMDs and the corresponding GPDs and TMDs are given in    \cite{Lorce13}.
For operator $\Gamma^{ij}= \delta^{ij}$, the gluon correlator is related to the  unpolarized gluon TMD $f_1^g$ and the gluon Sivers function $f_{1T}^{\perp g}$.
As we did not consider the  gauge link at light-cone infinity, we cannot access the gluon Sivers function in this model.  The GPD limit for this case gives $H_g$ and $E_g$. The gluon operator with $\Gamma^{ij}=- i\,\epsilon^{ij}$ is related to the gluon helicity TMD $g_{1L}^g$ and the worm gear function $g_{1T}^g$ (transverse helicity). The GPD limit of this operator gives $\tilde H_g$ and $\tilde E_g$.  In our case, $W_{LL}({\bs k _\perp}, {\bs b _\perp})$ is related to the  gluon helicity TMD and $W_{TL}({\bs k _\perp}, {\bs b _\perp})$ is related to the worm gear function. For $\Gamma^{ij}=\Gamma^{RR}$,  the TMD limit corresponds to four leading twist TMDs, gluon Boer-Mulders function $h_1^{\perp g}$, the other worm-gear function $h_{1L}^{\perp g}$ (longitudinal transversity), $h_{1T}^g$ which is part of gluonic transversity TMD and the gluon pretzelocity distribution, $h_{1T}^{\perp g}$. In our model, we cannot access the  pretzelocity distribution  and the worm-gear function, $h_{1L}^{\perp g}$, as they need the contribution of the gauge link.  $W_{U\T}({\bs k _\perp}, {\bs b _\perp})$ is related to one of the gluon Boer-Mulders functions  $h_1^{\perp g}$ \cite{Buffing13} , $W_{T\T}({\bs k _\perp}, {\bs b _\perp})$ is related to $h_{1T}^g$.  
The GPD limit of this operator structure is related to the chiral odd GPDs. 
\section{Gluon GTMDs and Spin Densities Relations}\label{density}
\begin{figure}[h]
(a)\includegraphics[width=7.5cm,height=5.5cm]{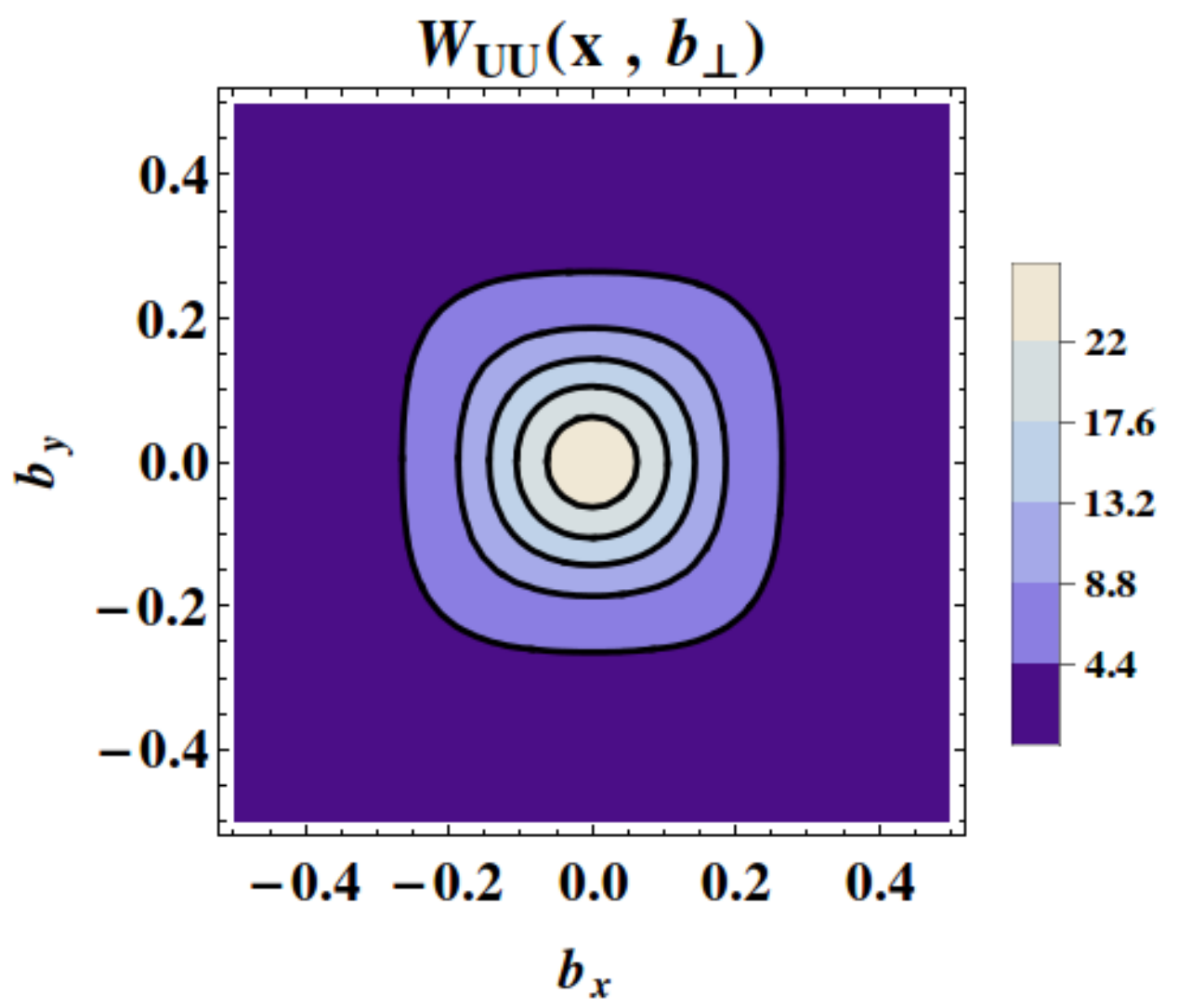}
(d)\includegraphics[width=7.5cm,height=5.5cm]{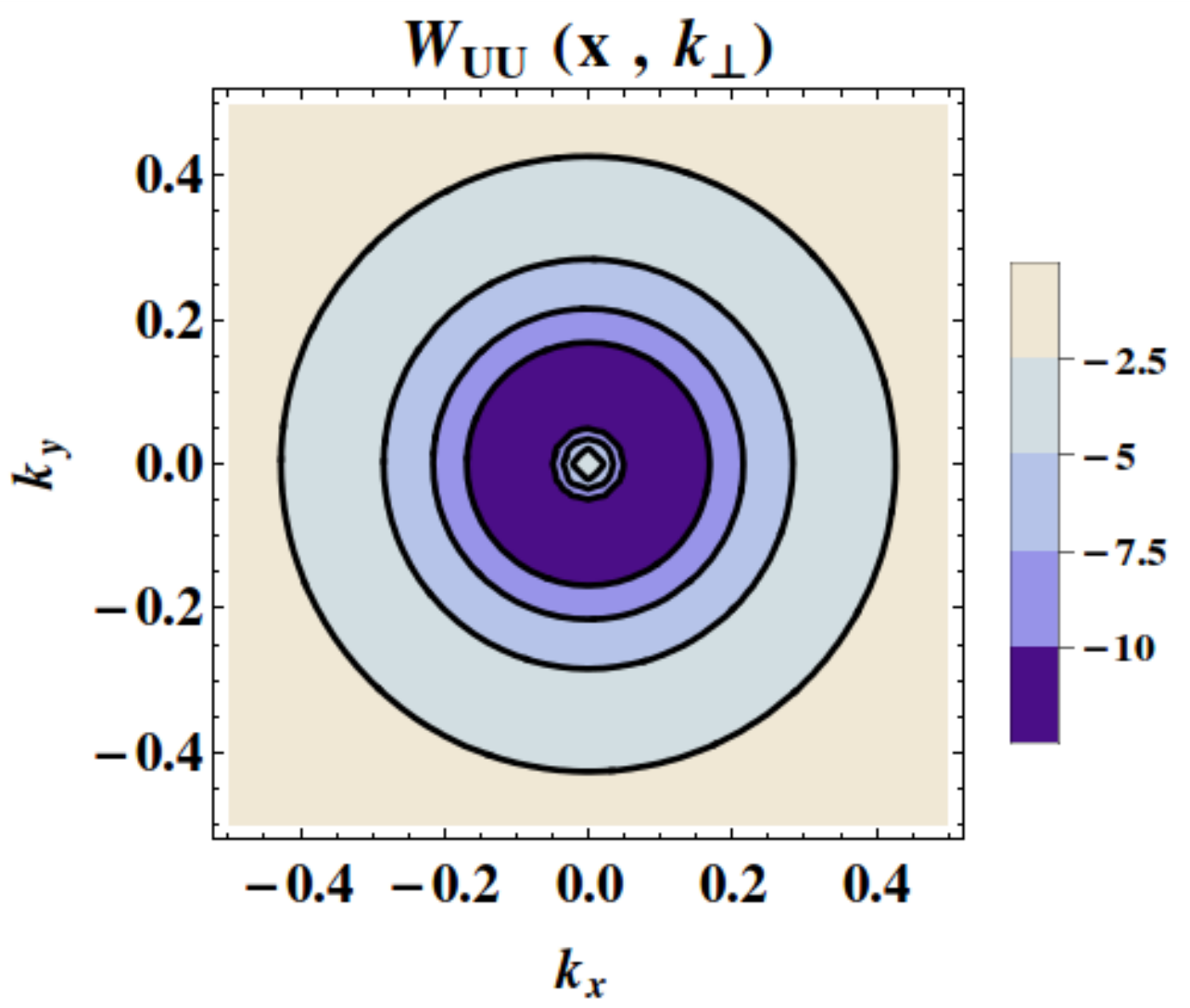}\\
(b)\includegraphics[width=7.5cm,height=5.5cm]{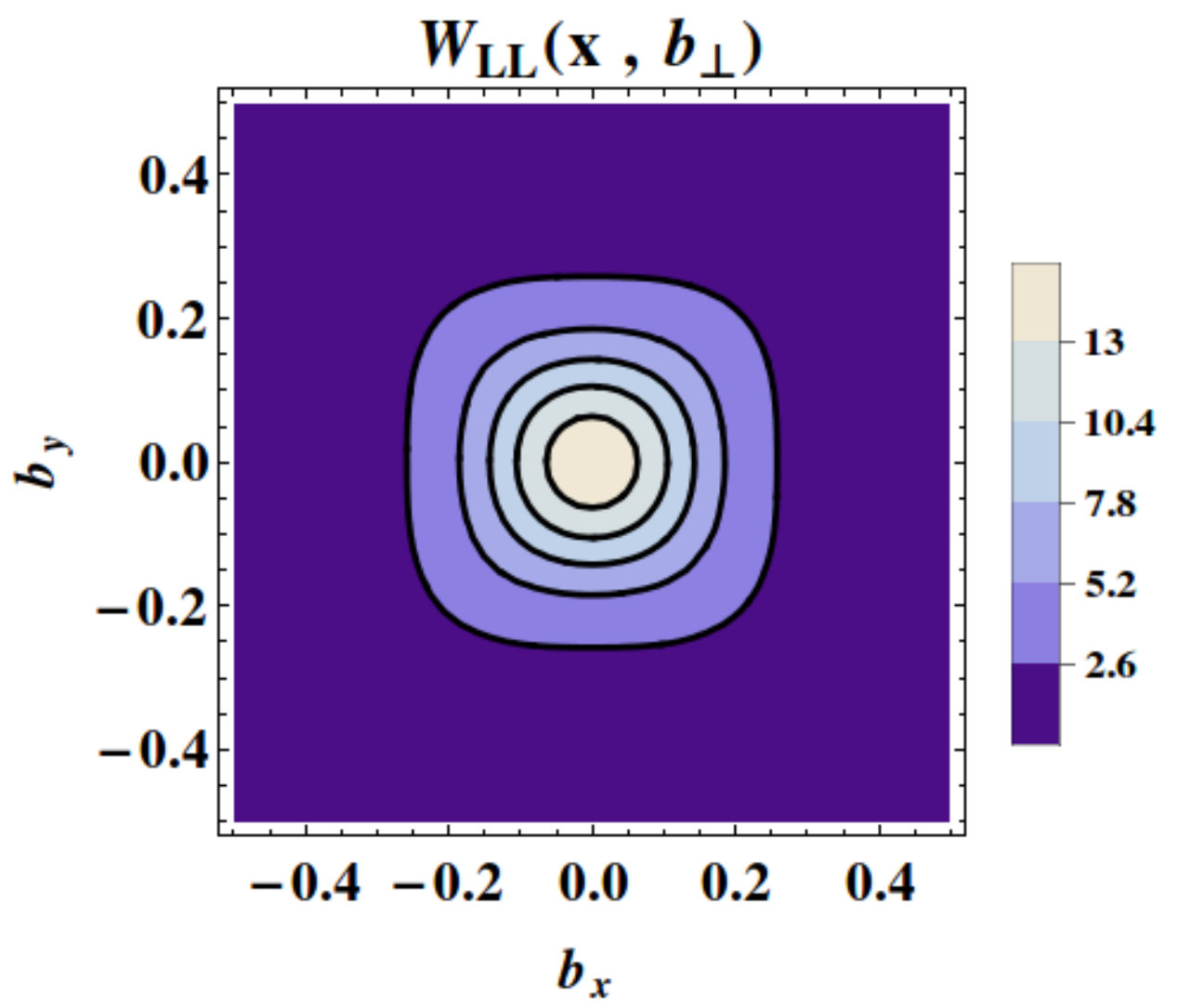}
(e)\includegraphics[width=7.5cm,height=5.5cm]{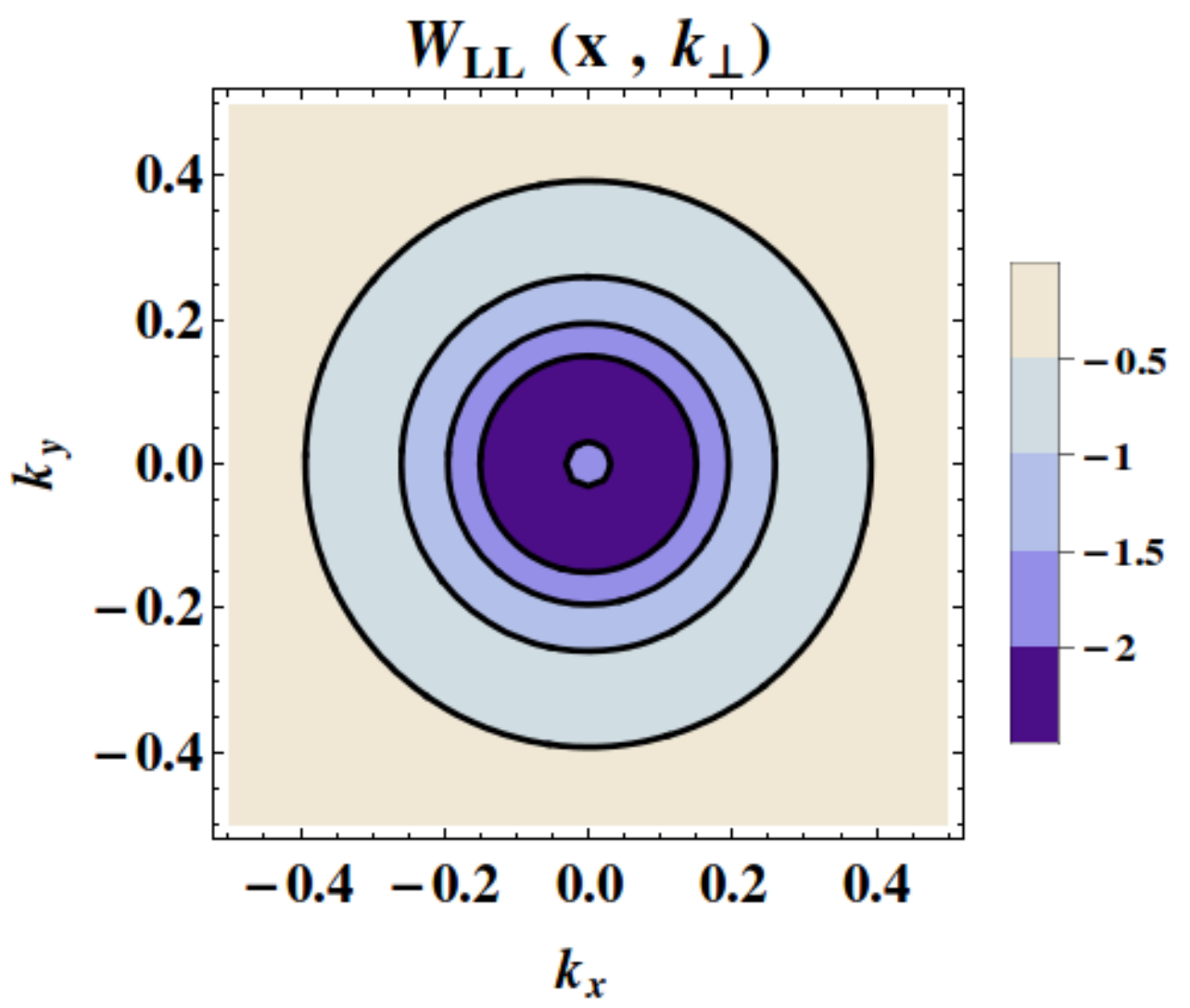}\\
(c)\includegraphics[width=7.5cm,height=5.5cm]{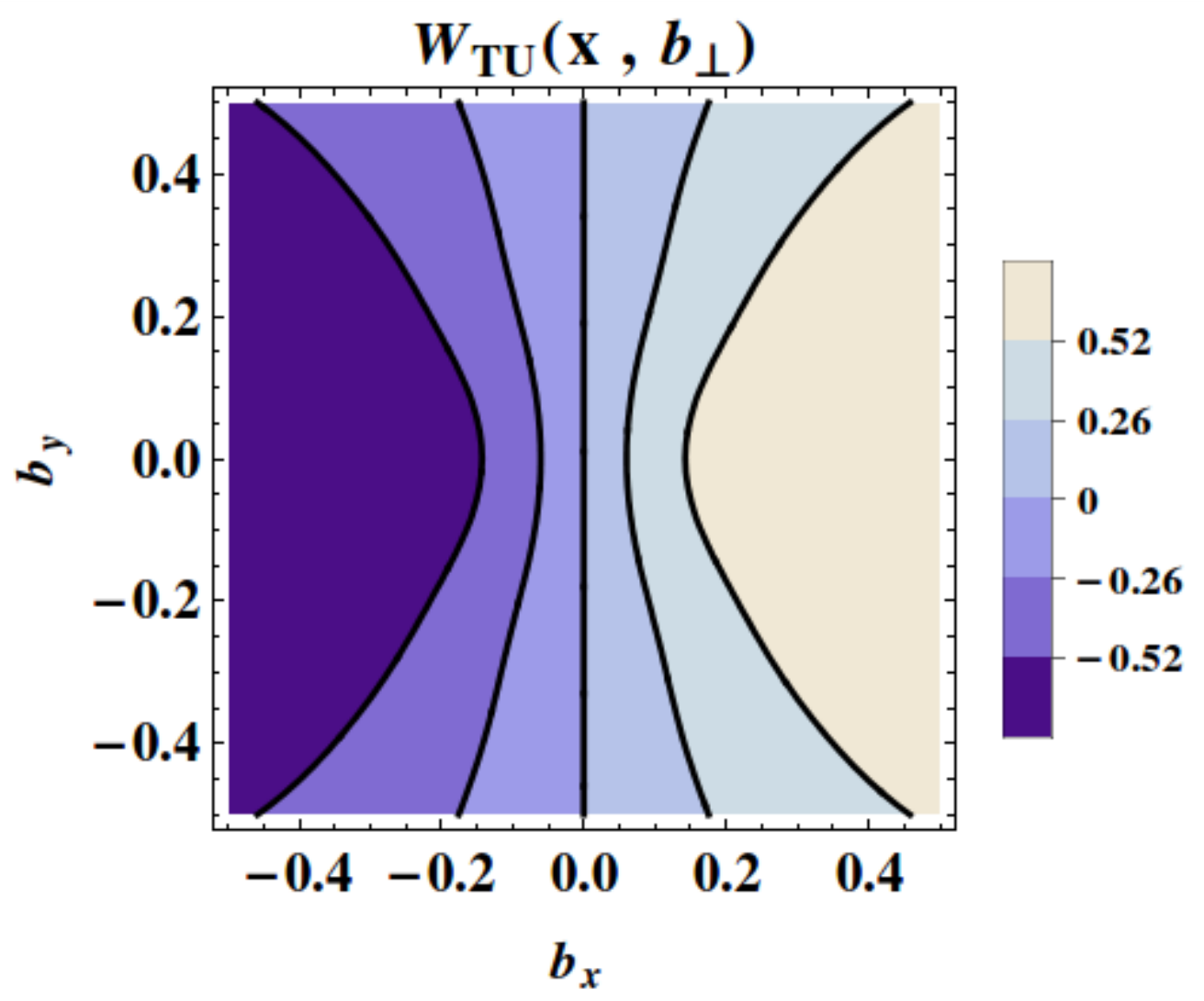}
(f)\includegraphics[width=7.5cm,height=5.5cm]{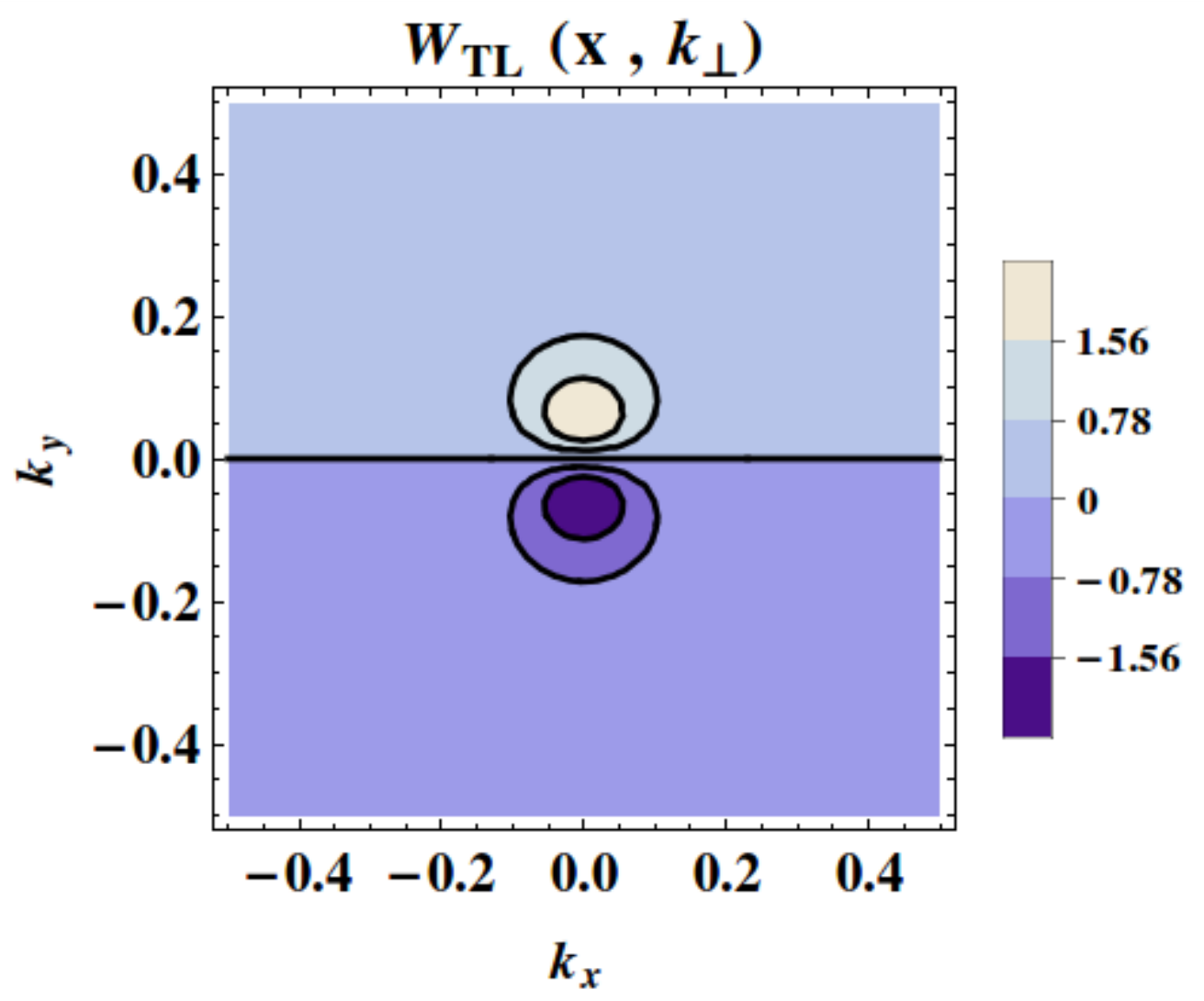}
  \caption{Left panel shows spin densities in impact parameter space for a fixed value of $x=0.3$. Right panel shows spin densities in momentum space for a fixed value of $x=0.3$}
 \label{tmd}
\end{figure}

The analytical expressions for gluon GTMDs for the unpolarized and longitudinally polarized gluon have been discussed in \cite{Mukherjee15}.  We will give here the expression for the new distribution that has been included in this work for the first time.
It should be noted here that we use the parameterization discussed in \cite{Meissner07} to extract the expression of GTMD given below.
\ba
 D^{2,+}_{1,1\,a}\es
 \frac{N}{D(q_\perp)D(q'_\perp)}\frac{m^2}{2\,k_1 k_2}
\frac{8 
\Big(k_1 k_2-\frac1{4}\Delta_1 \Delta_2 (1-x)^2\Big)}
{(1-x) x^3}
\label{Dfull}
\ea 
$k_1,k_2$ and $\Delta_1, \Delta_2$ are components of ${\bs k}_\perp$ and ${\bs \Delta}_\perp$ respectively. 
As discussed before in Eq. (\ref{wignerdistribution}),  Wigner distributions are defined as a two dimensional Fourier transformation of GTMDs. Similarly, transverse densities and densities in impact parameter space are two dimensional Fourier transform of Form Factors and GPDs respectively. Analogously we integrate Wigner distributions over impact parameter space and transverse momentum space to obtain spin densities in momentum and impact parameter space respectively. Spin densities are shown in Fig. \ref{tmd}. For this plot, we have taken a fixed value of $x=0.3$.  Spin density for unpolarized gluon in an unpolarized target state, as well as longitudinally polarized gluon in a longitudinally polarized target state, are peaked at the center. Spin density for unpolarized  gluon in a transversely polarized target is spread over the ${\bs b}_\perp$ space, and the spin density for longitudinally polarized  gluon in a transversely polarized target in ${\bs k}_\perp$ space is concentrated at the center.   
\subsection{TMD limit} 
TMDs can be  obtained from GTMDs by taking the forward limit.
In our model we set the gauge link to unity hence can acesses only T-even TMDs, so we display the relevant relations of gluon TMDs and GTMDs at leading twist \cite{Lorce13}. 
\ba
f_1^g(x,\, {\bs k}_\perp)\es \Re e [ S^{0\, +\,g}_{1,1\, a} (x,\, {\bs k}_\perp)]\label{S0p}\\
g_{1L}^g(x,\, {\bs k}_\perp)\es \Re e [ S^{0\, -\,g}_{1,1\, a}(x,\, {\bs k}_\perp)]\label{S0m}\\
g_{1T}^g(x,\, {\bs k}_\perp)\es \Re e [ P^{0\, -\,g}_{1,1\, a}(x,\, {\bs k}_\perp)]\label{P0m}\\
h_1^{\perp g}(x,\, {\bs k}_\perp)\es 2 \Re e [ D^{2\, +\,g}_{1,1\, a}(x,\, {\bs k}_\perp)]
\ea
The corresponding analytical expressions in our model are:
\ba
f_1^g(x,\, {\bs k}_\perp)\es \frac{2 N}{x}\,\,\frac{ k_\perp^2(x^2-2x+2) + m^2 x^4}{(k_\perp^2+m^2 x^2)^2}\label{f1}\\
g_{1L}^g(x,\, {\bs k}_\perp)\es2 N\,\,\frac{ k_\perp^2(2-x) + m^2 x^3}{(k_\perp^2+m^2 x^2)^2}  \label{g1l}\\
g_{1T}^g(x,\, {\bs k}_\perp)\es -4 N x(1-x)\,\,\frac{m^2}{(k_\perp^2+m^2 x^2)^2}\label{g1t} \\
h_1^{\perp g}(x,\, {\bs k}_\perp)
\es  8 N \frac{(1-x)}{x} \frac{m^2}{(k_\perp^2+m^2 x^2)^2}
\label{h1}
\ea
In extracting the TMDs and GPDs we have taken into account that our starting expression of the operator structure in the Wigner distribution has an overall negative sign compared to operator structure in \cite{Meissner07}. The expressions of the TMDs in Eq. (\ref{f1})-(\ref{h1}) match with Ref \cite{Meissner07}.
\subsection{GPDs in impact parameter space}
Impact parameter dependent distributions (IPDs) can be obtained by integrating the Wigner distributions over the transverse momentum. In our model calculation we have taken the skewness parameter $\xi$ to be zero. So the IPDs can be interpreted as the densities of gluons with longitudinal momentum fraction $x$ in the transverse position ${\bs b}_\perp$ with respect to center of momentum of the dressed quark.
We integrate the Wigner distributions over the  transverse momentum  and compare it with the parameterizations for gluon generalized correlators in \cite{Lorce13, Meissner07} to  obtain the IPDs. 
\ba
\mathcal{H}^g(x,\, {\bs b}_\perp)\es 2N \int [k \Delta]\,\,\, \frac{m^2 x^4+(x^2-2x+2) \Big({\bs k _\perp}^2-\frac1{4}{\bs \Delta _\perp}^2 (1-x)^2\Big)}{x\,\tilde{D}(x, {\bs k}_\perp, {\bs \Delta}_\perp)}\label{Hg}
\nn
\\[2ex]
 \mathcal{E}^g(x,\, {\bs b}_\perp)\es - 4N x(1-x)^2  \int [k \Delta]\,\,\,\frac{m^2}{\tilde{D}(x, {\bs k}_\perp, {\bs \Delta}_\perp)}\label{Eg}\\
\tilde{\mathcal{H}}^g(x,\, {\bs b}_\perp)\es 2N  \int [k \Delta]\,\,\, \frac{m^2 x^3+(2-x) \Big({\bs k _\perp}^2-\frac{1}{4}{\bs \Delta _\perp}^2 (1-x)^2\Big)}{\tilde{D}(x, {\bs k}_\perp, {\bs \Delta}_\perp)}\label{Htildeg}\\[2ex]
\mathcal{H}_T^g(x,\, {\bs b}_\perp)\es -4N x(1-x)  \int [k \Delta]\,\,\,\frac{m^2}{\tilde{D}(x, {\bs k}_\perp, {\bs \Delta}_\perp)}\label{HTg}\\
\mathcal{E}_T^g(x,\, {\bs b}_\perp)\es -16N \frac{(1-x)}{x}\int [k \Delta]\,\,\,\frac{\frac{m^2 }{\Delta_1 \Delta_2} (k_1 k_2-\frac{1}4(1-x)^2\Delta_1 \Delta_2)}{ \tilde{D}(x, {\bs k}_\perp, {\bs \Delta}_\perp)}\label{ETg}\\
\tilde{\mathcal{H}}^g_T(x,\, {\bs b}_\perp)\es 0\label{HTtildeg}
\ea
where
\ba
\int [k \Delta]\es\int  d^2 {\bs k}_\perp\int \frac{d^2 {\bs \Delta}_\perp \cos({\bs \Delta _\perp}{\bs b _\perp})}{2(2 \pi)^2} \\
\tilde{D}(x, {\bs k}_\perp, {\bs \Delta}_\perp)\es\left(({\bs k}_\perp+\frac{{\bs \Delta}_\perp(1-x)}{2})^2+m^2x^2\right)\left(({\bs k}_\perp-\frac{{\bs \Delta}_\perp(1-x)}{2})^2+m^2x^2\right)\nn 
\ea
Thus, in our model $\tilde{\mathcal{E}}^g$ and $\tilde{\mathcal{E}}^g_T$ do not contribute since we have chosen $\xi=0$.
The relations we obtain in Eq. (\ref{Hg}) -- (\ref{HTtildeg}) are in agreement with the GPDs obtained in \cite{Meissner07}.
\section{Quark and Gluon Helicity and Orbital Angular Momentum at small-$x$}\label{smallx}
The behavior of the GTMDs and TMDs at small $x$ has attracted quite a lot of interest recently. In \cite{Piet17} a bound on the gluon TMDs at small $x$ was studied using dipole type gauge link.  As stated before, the GTMDs are related to the quark and gluon helicity distributions and OAM. The latest extraction of gluon helicity distribution $\Delta G(x) $ at small $x$ suffers from large uncertainties, and in this context a good estimate of the quark and gluon helicity and OAM in the small  $x$ region  are essential as they all contribute to the nucleon spin. Several authors recently explored the quark and gluon GTMDs in this kinematical region. In \cite{Hatta16} a relation between gluon helicity and OAM at small $x$ was introduced and further discussed in \cite{Hatta18}. Quark helicity distribution and OAM were  also investigated in \cite{Hatta18} in two different models. The scale evolution of the different contributions to the nucleon spin  was also studied, with special emphasis in the small $x$ region, which was not done in previous studies \cite{HandK}. As is well known, a dressed quark  in QCD or a dressed electron in QED provides a very good model for an intuitive description of the spin 
and angular momentum of a composite relativistic system \cite{Brodsky01}. In this section, we present a few observations and discussions about the quark and gluon helicity and OAM distributions in the dressed quark model  in the limit of small $x$. Here we consider the Jaffe-Manohar sum rule for the spin:
\ba
{1\over 2} = {1\over 2} \Delta \Sigma + l_q + \Delta G +l_g
\label{sumrule}
\ea
For a nucleon, ${1\over 2} \Delta \Sigma $ is the contribution from intrinsic spin of the quarks (and antiquarks), $\Delta G$ is the intrinsic spin of the gluons, $l_q$  and $l_g$ are the canonical quark and gluon OAM, respectively. The terms on the right hand side of the above equation are the respective distributions integrated over $x$. The quark and gluon helicity and OAM are directly extracted from the GTMDs \cite{Mukherjee14, Mukherjee15}. For a dressed quark target, the leading log contributions are obtained in the massless limit of the quark \cite{HandK}. Our results as given below agree with \cite{HandK} in the massless limit.     

The canonical OAM distribution for the quark is obtained from the GTMD  \cite{Lorce11,Hatta12,Lorce12},
\ba
l_q(x)=-\int d^2 {\bs k}_\perp {{\bs k}_\perp^2 \over m^2}\,\, S^{0\, +q}_{1,1\, b}(x, {\bs k}_\perp)
\ea
Quark helicity is obtained from the  GTMD,  
\ba
{1\over 2} \Delta \Sigma(x) = \int d^2{\bs k _\perp} S^{0\, -q}_{1,1\, a}(x,{\bs k _\perp})
\label{rel2}
\ea

The canonical OAM distribution for the gluon is obtained from the GTMD  \cite{Lorce11,Hatta12},
\ba
l_g(x)=-\int d^2 {\bs k}_\perp {{\bs k}_\perp^2 \over m^2}\,\, S^{0\, +g}_{1,1\, b}(x, {\bs k}_\perp)
\ea
Gluon helicity is obtained from the  GTMD,  
\ba
\Delta G(x) =\int d^2{\bs k _\perp} S^{0\, -g}_{1,1\, a}(x,{\bs k _\perp})
\label{rel2}
\ea

By comparing the operator structures in the above two cases for small $x$,  the following relation between the gluon OAM and gluon helicity was pointed out by Hatta {\it et.al.} \cite{Hatta16} (note that the $l_g$ used in this reference has a difference in a factor of 2 compared to ours, as seen from Eq. (32) in Ref \cite{Hatta16}, however, our results agree with \cite{Hatta18}); 
\ba
l_g(x) \approx -\Delta G(x)
\label{rel1}
\ea

That is the gluon helicity distribution at small $x$  cancels the contribution from gluon OAM. The relation above was obtained in \cite{Hatta16} using an operator analysis without any reference to the small $x$ region. In \cite{Hatta18} this relation is shown to emerge from a Regge-like behaviour of $\Delta G(x)$ and $l_g(x)$ at small $x$. Note that here $x$ is the momentum fraction of the gluon.
 
The gluon GTMDs in the dressed quark model was evaluated in Ref.~\cite{Mukherjee15} at the leading twist.
The gluon canonical orbital angular momentum $l_g(x)$ was also calculated using these GTMDs. Because of a negative sign in the operator structure of our starting expression of the Wigner distribution as compared to \cite{HandK}, we need an extra negative sign on the right-hand side of Eq. (\ref{rel2}), in order to compare with \cite{HandK} in the massless limit. 
The analytic expressions for $l_g(x)$ and $\Delta G(x)$ are given by
\ba
l_g(x) \es - {\alpha\over 2 \pi} C_F (1-x)(2-x)\,\left[I_g(x) - \pi \,\right]\\[2ex]
\Delta G(x) \es -{\alpha\over 2 \pi} C_F \left[
(x-2)\,I_g(x)+2\pi\,(1-x)
\right]
\label{rel3} \ea
where
\ba
I_g(x) = \pi\, \log \left(
\frac{Q^2+m^2\,x^2}{\mu^2+m^2\,x^2}\right)
\ea
Here $Q$ and $\mu$ are the upper and lower cutoff on ${\bs k_\perp}$ and $\alpha= {g^2 \over 4 \pi}$. From the analytic expression above, it is clear that
the relation (\ref{rel1}) holds in our model for small $x$.  
For completeness, we also investigate the quark helicity and OAM distributions at small $\tilde x=1-x$, where $\tilde x$ is the momentum fraction of the quark in our model. These can be calculated using the GTMD expressions calculated before \cite{Mukherjee14}. The analytic expressions for $l_q(\tilde x)$ and ${1\over 2} \Delta \Sigma(\tilde x)$ are given by
\ba
l_q(\tilde x)\es - {\alpha\over 2 \pi} C_F (1-\tilde x^2)(I_q(\tilde x)-\pi)\nn
{1\over 2} \Delta \Sigma(\tilde x)\es {\alpha \over 2 \pi} C_F [I_q(\tilde x) \frac{(1+\tilde x^2)}{(1-\tilde x)}-2 \pi\frac{(1-\tilde x+\tilde x^2)}{1-\tilde x}]\nn 
\ea 
where $$I_q(\tilde x)=I_g(1-x)$$
\begin{figure}[h!] 
 \centering 
\includegraphics[width=8cm,height=5.5cm]{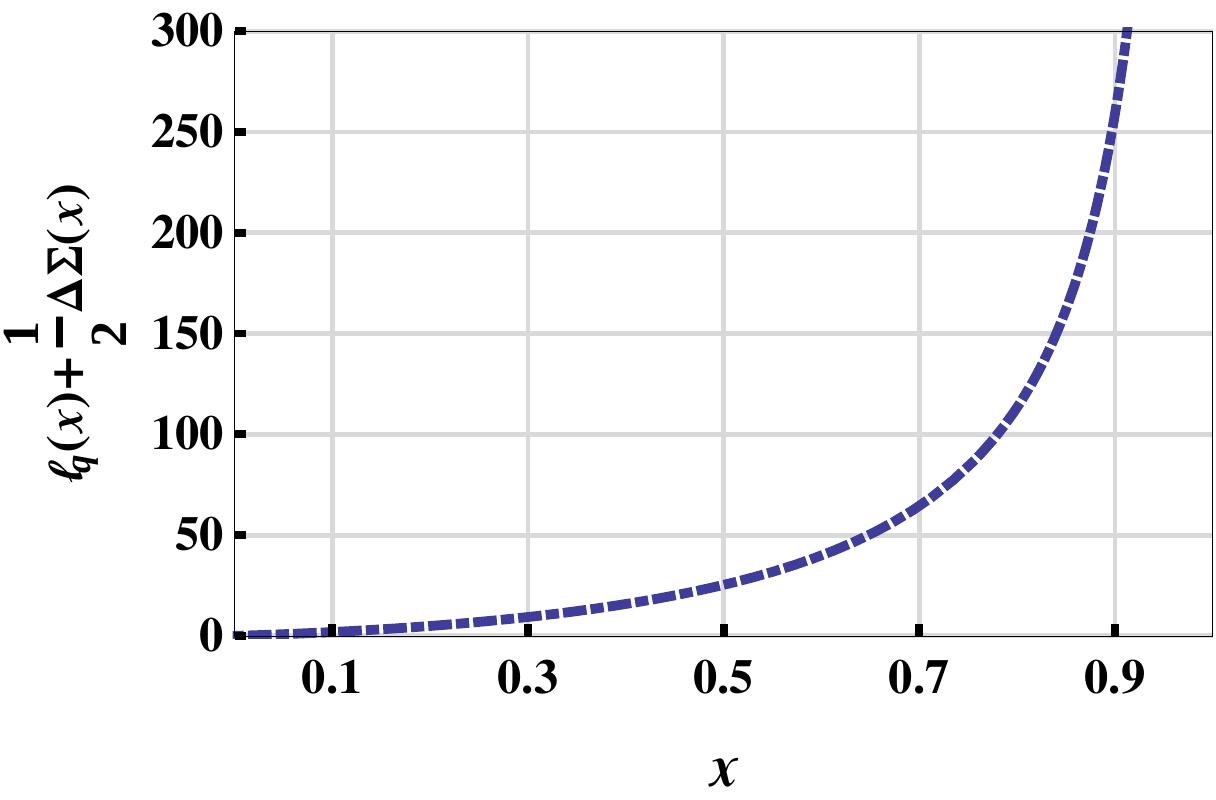}
\caption{Plot of  $l_q(x)+{1\over 2 }\Delta \Sigma (x)$ vs $x $. We have taken ${\alpha\over 2 \pi} C_F =1$. We  have chosen  $Q=10~\mathrm{GeV}$ and  $\mu=0.0~\mathrm{GeV}$.}  \label{relplot1}
\end{figure}
The expressions for  $l_g(x)$, $l_q(x)$ and $\Delta G(x)$ as well as $\Delta \Sigma(x) $ given above argree with  \cite{HandK} in the  massless limit of the quark.  We also verify that the  following relation holds for small $\tilde x$  
\ba
l_q(\tilde x) \approx - {1\over 2} \Delta \Sigma(\tilde x)
\label{rel5} 
\ea

Both Eqs. (\ref{rel1}) and (\ref{rel5}) become exact equalities  in the massless limit of the quark.  
That is when the momentum fraction of the quark(gluon)  probed is small, the quark(gluon)  helicity distribution in this model cancels the  quark(gluon)  OAM distribution, similar to the behavior predicted in \cite{Hatta18} for quark and gluon distributions of the nucleon at small $x$. It is to be noted that all the four distributions, quark and gluon, are non-zero in the small $x$ limit. In order to satisfy the sum rule, it is important to take into account the contribution from the single particle sector of the Fock  space to $\Delta \Sigma (\tilde x)$, and the normalization of the state \cite{HandK}.    

\begin{figure}[h!] 
 \centering 
\includegraphics[width=8cm,height=5.5cm]{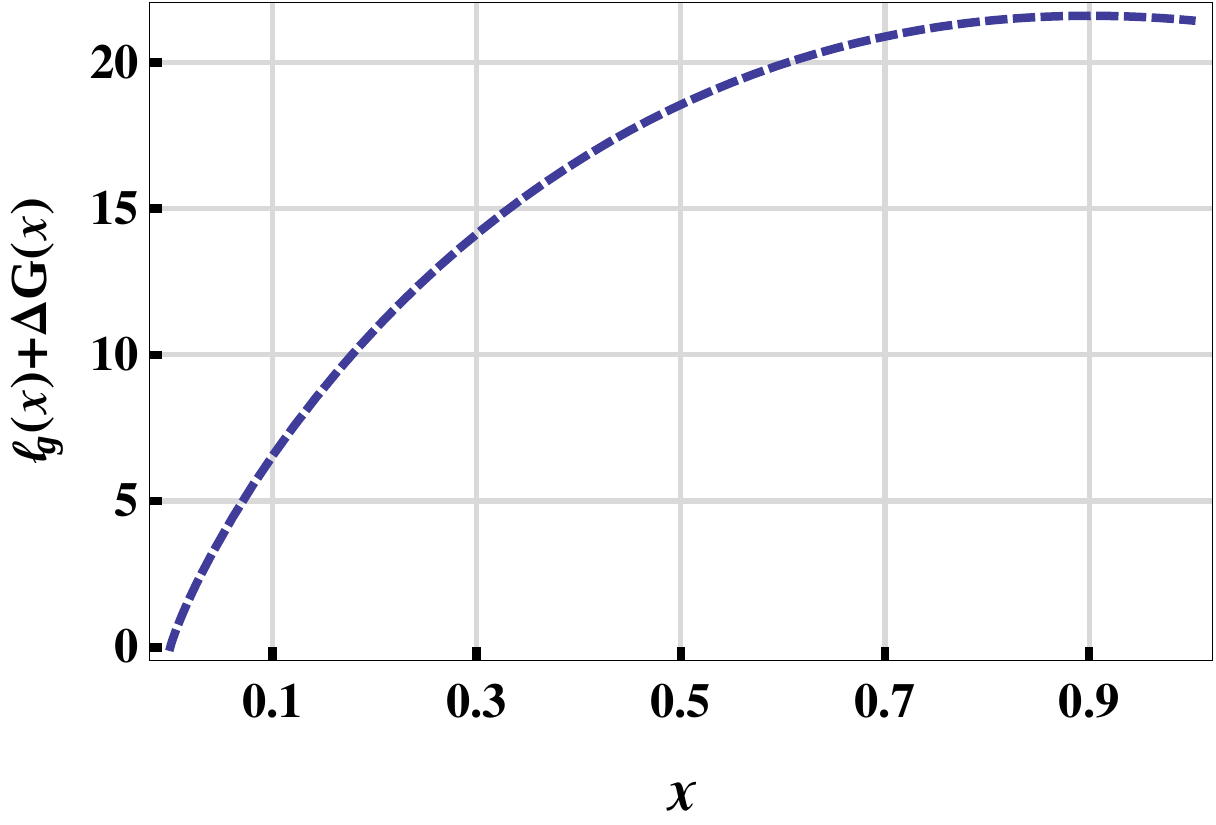}
  \caption{Plot of $l_g(x)+\Delta G(x)$ vs $x$. We have chosen $Q=10~\mathrm{GeV}$, $\mu=0.0~\mathrm{GeV}$ and ${\alpha\over 2 \pi} C_F =1$. }  \label{relplot2}
    
\end{figure}
\begin{figure}[h!] 
 \centering 
\includegraphics[width=8cm,height=6cm]{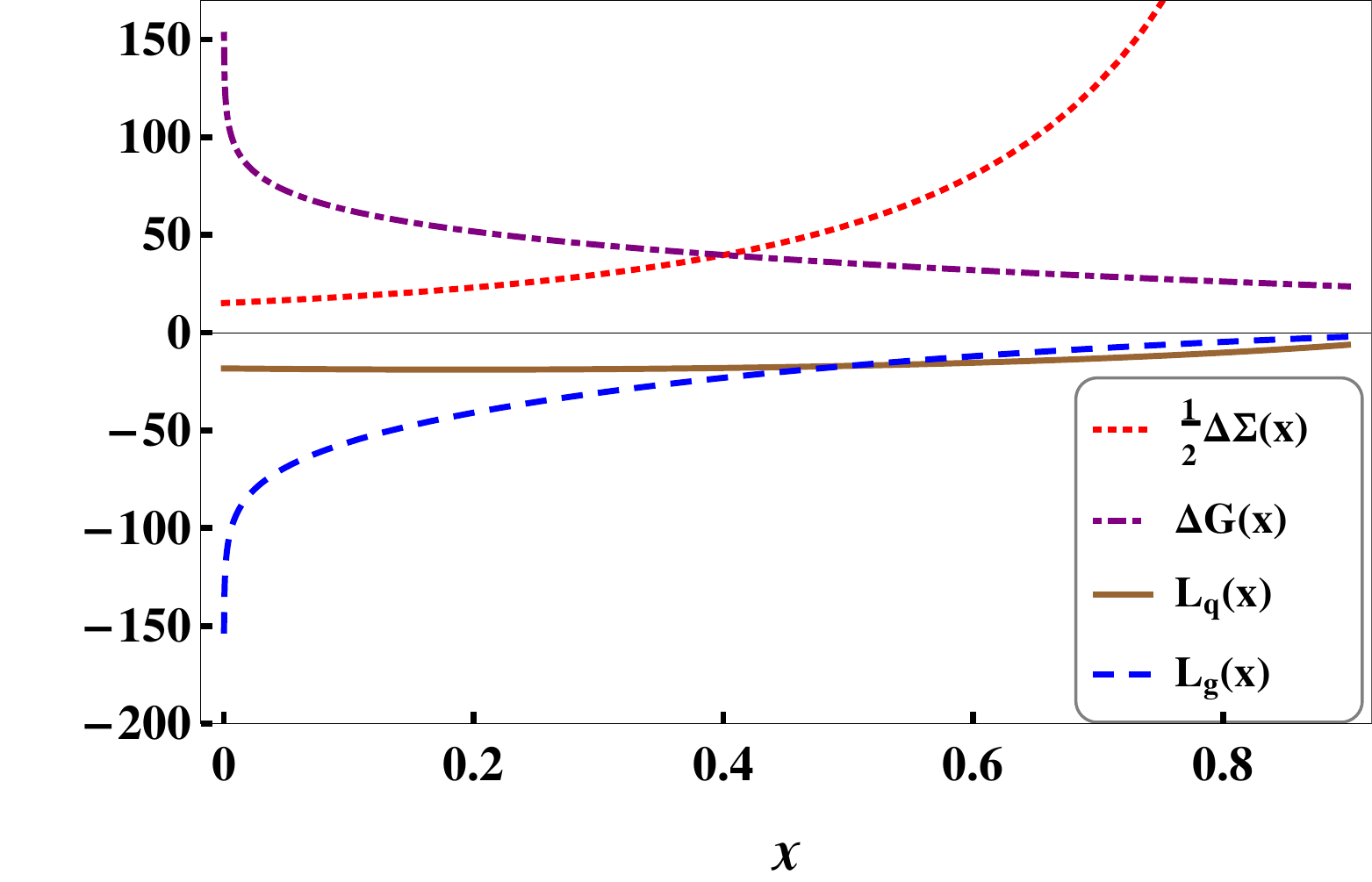}
  \caption{ Plot of the quark and gluon helicity distributions and OAM distributions vs $x$. We have taken ${\alpha\over 2 \pi} C_F =1$. We  have chosen  $Q=10~\mathrm{GeV}$ and  $\mu=0.0~\mathrm{GeV}$. }  \label{relplot3}
\end{figure}

Figures~\ref{relplot1} and \ref{relplot2}  show  the  small-$x$ relation given by Eqs. (\ref{rel5}) and (\ref{rel1}) respectively. The small non-zero value at small $x$ is obtained from the quark mass term. Note that we have taken the mass of the target and the probed quark to be the same.  In Fig. \ref{relplot3} we have shown all the four distributions as functions of $x$, where $x$ is the momentum fraction of the quark/gluon probed. The gluon OAM and helicity distributions need a regulator $\mu$ as $x \to 0$, as seen in the analytic expression. In some versions of this model, for example in the corresponding QED model,  a non-zero mass is introduced  for the internal gauge boson to regulate this divergence \cite{Brodsky01}. However, the divergent logarithmic  piece cancel in the sum of $l_g(x)$ and $\Delta G(x)$ for small $x$. 

\section{Conclusion}\label{conclusion}   
In this work, we have investigated the gluon Wigner distributions for unpolarized, longitudinally polarized and transversely polarized target state, taking into account all possible polarization combinations of the gluon at leading twist.  Instead of a proton target, we took a quark state dressed with a gluon, which acted as a perturbative model of a composite spin-$1/2$ state, and made it possible to investigate the gluon Wigner distributions, unlike most phenomenological models without a gluonic degree of freedom. 
 We improved the numerical convergence compared to an earlier study with respect to the upper limit on $\Delta_\perp (\Delta_{max})$, as a result, the Wigner
 distributions presented here are independent of $\Delta_{max}$. We study the leading twist gluon Wigner distributions, out of which 6 
are independent Wigner distributions, and the remaining are  linear combinations of those. 
 All the distributions are expressed as overlaps of light-front wave functions and are calculated using an analytic form of the wave functions in light-cone gauge,
 they are presented in the transverse position, transverse momentum and mixed space.  We have chosen light-cone gauge and have taken
the gauge link to be unity. We obtained the GTMDs and calculated the spin densities in transverse momentum and position space. 
We have investigated the quark and gluon helicity and OAM distributions at small $x$ and shown that in this model the quark(gluon) helicity is largely canceled  by the quark(gluon) OAM at small-$x$, a behavior  that is already proposed earlier in the literature. 
 Further work in this line would be to investigate the effect of the gauge link on the Wigner distributions and TMDs.

\end{document}